\documentclass[iop]{emulateapj}

\usepackage{natbib}
\journalinfo{Astrophysical Journal}
\slugcomment{Received Sept 7, 2015 ; Accepted Dec 15, 2015}

\def\nab{\mbox{\boldmath $\nabla$}}

\providecommand{\e}[1]{\ensuremath{\times 10^{#1}}}
\newcommand\gsim{\,\lower3pt\hbox{$\sim$}\llap{\raise2pt\hbox{$>$}}\,}
\newcommand\lsim{\,\lower3pt\hbox{$\sim$}\llap{\raise2pt\hbox{$<$}}\,}

\begin{document}

\title{Theoretical limits on magnetic field strengths in low-mass stars}

\author{Matthew K. Browning\altaffilmark{1}, Maria A. Weber\altaffilmark{1},
  Gilles Chabrier\altaffilmark{1,2}, Angela P. Massey\altaffilmark{3}}


\altaffiltext{1}{Dept of Physics and Astronomy, Stocker Road, University of Exeter, EX4 4QL,
  browning@astro.ex.ac.uk}

\altaffiltext{2}{Ecole Normale Superieure de Lyon, CRAL, UMR CNRS 5574, F-69364 Lyon Cedex 07, France}

\altaffiltext{3}{Department of Astronomy, Boston University, 725 Commonwealth Avenue, Boston, MA 02215, USA}

\begin{abstract}

  Observations have suggested that some low-mass stars have larger radii
  than predicted by 1-D structure models.  Some theoretical models have
  invoked very strong interior magnetic fields (of order 1 MG or more) as a
  possible cause of such large radii.  Whether fields of that strength
  could in principle by generated by dynamo action in these objects is
  unclear, and we do not address the matter directly.  Instead, we examine
  whether such fields could remain in the interior of a low-mass object for
  a significant time, and whether they would have any other obvious
  signatures.  First, we estimate timescales for the loss of strong fields
  by magnetic buoyancy instabilities.  We consider a range of field
  strengths and simple morphologies, including both idealized flux tubes
  and smooth layers of field.  We confirm some of our analytical estimates using
  thin flux tube magnetohydrodynamic (MHD) simulations of the rise of
  buoyant fields in a fully-convective M-dwarf. Separately, we consider the
  Ohmic dissipation of such fields.  We find that dissipation provides a
  complementary constraint to buoyancy: while small-scale, fibril fields
  might be regenerated faster than they rise, the dissipative heating
  associated with such fields would in some cases greatly exceed the
  luminosity of the star.  We show how these constraints combine to yield
  limits on the internal field strength and morphology in low-mass stars.
  In particular, we find that for stars of 0.3 solar masses, no fields in
  flux tubes stronger than about 800 kG are simultaneously consistent with
  both constraints.  
  
\end{abstract}


\section{INTRODUCTION}

Low-mass stars are magnetic.  In many cases that magnetism is observed at
the stellar surface, either through proxies like chromospheric activity or
coronal emission \citep[e.g.,][]{noyes_ea84, Pizzolato_etal2003, wright_etal2011},
through direct photometric monitoring of starspots
\citep[e.g.][]{basri_walkowicz_reiners2013,mcquillan_etal2014}, or through
measurements of Zeeman broadening in magnetically sensitive lines or
molecular bands \citep[e.g.,][]{reiners07}.  Stellar flares of remarkable
strength are also commonplace in low-mass stars
\citep[e.g.,][]{2010apj...721..785o} and are similarly believed to be
linked to the presence and properties of a strong magnetic field.

The magnetism of these stars may influence their structure and evolution in
a variety of ways.  One of the most remarkable clues that this might be
occurring has come from studies of the radii of active, low mass stars: in
eclipsing K-dwarf and M-dwarf binaries, measured radii exceed model
predictions in some cases by about 10 per cent \citep[e.g.,][]{torres2002,
  lopezmorales2007, mor08, stassun_etal2012, torres2013}. Most of the
  systems for which precise radius measurements are available are in close
  eclipsing binaries, which are expected to be rapidly rotating and
  strongly magnetic, though discrepancies also appear to be present in at
  least one longer-period system \citep{irwin_etal2011_lspm}.  Several
  authors have attributed these seemingly anomalous radii and temperatures
  to the presence of remarkably strong interior magnetic fields.
  \citet{mullan2001}, for example, argued that some of the observed
  properties of these stars could be explained if the interior was not
  fully convective but instead possessed a magnetic field strong enough to
  render the innermost parts of the star stable to convection.  Using the
  stability analysis of \citet{gough_tayler1966}, they estimated the
  required field strengths to be as great as 100 MG.  Many other works have
  argued that less extreme magnetic fields might still lead to noticeable
  changes in the stellar radius, either through inhibition of convective
  transport or through the effects of cool starspots
  \citep[e.g.,][hereafter collectively referred to as MM]{spruit_weiss1986, chabrier2007, macdonald_mullan2012,
    macdonald_mullan2013, macdonald_mullan2014, macdonald_mullan2015}. Very
  recently, \citet[][hereafter FC14]{feiden_chaboyer2012,
    feiden_chaboyer2013,feiden_chaboyer2014} have presented stellar models
  using a modified version of the Dartmouth stellar evolution code that
  attempts to account for the effects of magnetism within the context of
  mixing-length theory; they concluded that only fields with strengths
  greater than 10 MG in the deep interior would be able to modify 
  convection enough to explain the observed radii of fully convective
  stars (though, as noted below, they ultimately found such fields untenable).

Whether fields of such strength could in reality be generated in a star, or
even just maintained if somehow produced, is not entirely clear. In general
the fields are believed to arise from the action of a magnetic dynamo -- a
process that converts kinetic energy to magnetic \citep[see,
  e.g.,][]{moffatt1978}.  The maximum field strengths achievable in
principle by dynamo action in any given instance are not easy to estimate:
saturation of the dynamo at a given field strength results when the
production of the field (through induction by the motion of the
electrically conducting fluids) balances its Ohmic dissipation, but both of
these depend sensitively on the properties of the flows and fields and for
realistic flows it is not generally possible to predict {\sl a priori} how
this balance will be achieved.  In many astrophysical contexts, a commonly
employed estimate is that the magnetic energy will reach equipartition with
the convective kinetic energy (or with whatever flow is responsible for
maintaining the field) -- see, e.g., \citet{cattaneo1991}, \citet{browning2008},
  \citet{featherstone_etal2009} for example discussions in the context of stellar
dynamos, \citet{passot_etal1995} in the context of turbulence in the
interstellar medium, \citet{kulsrud_etal1997} in relation to galactic
magnetism, or \citet{thompson_duncan1993} in regards to magnetars.  This is
motivated in some cases by the fact that in a closed system with no
dissipation, the sum of magnetic and kinetic energies is identically
conserved, so any growing field must come at the expense of the flows; in
these systems the initial kinetic energy of the flows provides a firm upper
bound on the achievable field strengths.  But stellar dynamos are not
closed systems, and the kinetic energy of the flows may be continually
replenished by the vast reservoirs of potential energy present in the
system.  It is also possible to arrive at the equipartition estimate by
considering plausible balances of forces.  For example, if the Lorentz
force $ {\bf j} \times {\bf B}$ (involving the magnetic field ${\bf B}$ and
the current density ${\bf j} = (c/4\pi) \nab \times {\bf B}$) is
approximated simply by $jB \sim B^{2}/L$ and balances inertial terms, which
are taken to scale as $\rho v \cdot \nabla v \sim \rho v^2 / L$, we recover
$B^2 \sim \rho v^2$ \citep[see, e.g.,][]{roberts2009}. But other seemingly
plausible force balances (e.g., of Lorentz and Coriolis forces) can yield
estimates of the field that are much greater than the field in
equipartition with convection.  Indeed, there are many examples of dynamos
in which the magnetic energy is believed to be vastly in excess of the
convective kinetic energy; the most well-known example is the Earth, where
the magnetic energy may exceed the convective energy by factors of 100 or
more \citep[e.g.,][]{roberts_king2013, stelzer_jackson2013}.  Thus, even some of the very strong
fields considered in, e.g., \citet{macdonald_mullan2015}, which possess energy densities that are far greater than the energy in convective motions (but less than the rotational or thermal energies), even if in our
opinion unlikely, are difficult to rule out conclusively on the basis of dynamo theory alone.

Here, we set aside the difficult question of how such fields might be
built, and assess instead whether they could survive for extended times and
hence have any observable impact.  We examine {\sl magnetic buoyancy} and
{\sl Ohmic dissipation} as complementary constraints on the field
distribution in these stars.    One of our basic goals is to determine whether fields of the
strength envisioned in some models ($> 10^6$ G), however generated, could
persist in stellar interiors.  More generally, although we are motivated partly by the observation of possible radius inflation in these stars, we aim to explore what limits can be placed on internal field strength without recourse to arguments about equipartition or the specifics of dynamo action in a particular object. We are not the first authors to consider the problem of low-mass star magnetism in this light: some of the possible constraints arising from magnetic buoyancy, for example, were explored by FC14 and \citet{mullan2001}.  In particular, FC14 noted that simple estimates of the buoyant properties of strong fields suggest that simple flux tubes of $>10^6$ G would need to have very small cross-sectional radii to avoid rapid buoyant rise; if they did rise, they would plausibly traverse the stellar interior on timescales of days to weeks.  For this and other reasons, FC14 ultimately concluded that such strong field strengths were unlikely.  Our work here serves partly to place this analysis on firmer ground, both by examining what scales of field are incompatible with the constraints of dissipative heating, and by comparing the rapid buoyant rise time to the (in some cases also quite rapid) regeneration times that are possible on small scales.

In \S 2 we briefly review the magnetic buoyancy of isolated flux tubes, the
instability of tubes initially in equilibrium, and the breakup of magnetic
layers through buoyancy instabilities.  For some very simple field
configurations, we give estimates of the time it would take for fields to
rise to the stellar surface through the action of such buoyancy
instabilities; we also assess under what circumstances a more general field
configuration might be stable to these processes.  We compare these to
order-of-magnitude estimates of the fastest timescales on which the field
might plausibly be regenerated to derive limits on the field strength at a
given spatial scale.  Because smaller-scale fields tend to be less
susceptible to buoyant rise, we also calculate for some specific field
strengths (e.g., those explored in the MM and FC2014 models) the largest
possible spatial scale the magnetism could have in a steady state.  In \S
3 we briefly verify some of our analytical estimates by turning to MHD
simulations of the rise of thin flux-tubes.  We confirm the basic results
of \S 2, specifically that smaller tubes rise slower and that very strong,
larger-scale fields rise on remarkably short timescales.

In \S 4 we consider complementary constraints arising from the Ohmic
dissipation of the fields.  We estimate the total power arising from the
dissipation of the fields considered in \S 2, which again depends on both
the strength of the field and its spatial structure.  In extreme cases the
dissipated energy would greatly exceed the total stellar luminosity, so we conclude
that these combinations of strength and morphology are unrealistic.  We use
this to draw some tentative conclusions about the maximum possible field
strengths achievable in stars and sub-stellar objects in \S 5, and regard
Figure \ref{fig:combinedconstraints} there as the main result of this
paper. We briefly discuss some possible extensions of this analysis to other masses in \S 6, and close in
\S 7 with a summary of our work and its main limitations.

\section{MAGNETIC BUOYANCY AS A LIMIT ON FIELD STRENGTHS AND MORPHOLOGIES}

Magnetic fields exert a force, and this force is often decomposed into
terms corresponding to a magnetic pressure ($B^2/8\pi$) and to tension
along the fieldlines (${\bf B} \cdot \nabla {\bf B}$).  (The latter is not
precisely analogous to tension on a string -- see, e.g., \citet{kulsrud2005} --
but the identification of these terms with tension and with pressure is
commonplace and we will employ it here.)  Because the fields exert a
pressure, magnetized gas in total pressure equilibrium with unmagnetized (or less
magnetized) surroundings will in general have a different density and/or
temperature than those surroundings.  These density or temperature
differences can in many instances lead to instability, whether the field is
composed of discrete structures or is instead continuously distributed.
The resulting magnetic buoyancy instabilities have been studied for decades
both analytically and using numerical simulations.  In this section we
briefly review the extensive literature on these instabilities, and assess
their possible relevance for the fields examined in, e.g., the FC2014 and
MM models.  We draw extensively on the reviews by
\citet{hughes_proctor1988}, \citet{hughes2007}, \citet{fan2009} and \citet{cheung_isobe2014}
in what follows. Readers who are already familiar with the buoyancy instability of both isolated flux tubes and smooth layers could proceed to \S2.3-2.4, where we use these concepts to constrain possible field strengths and morphologies in low-mass stars.

Arguably the simplest configuration that exhibits magnetic buoyancy is an isolated,
magnetized ``flux tube,'' considered to be in pressure and thermal
equilibrium with its field-free surroundings \citep{parker1955, parker1975}. Such
structures will have a density deficit relative to their surroundings (as
demonstrated in more detail below), and will therefore tend to rise.  As
the tube rises, it will be buffeted by convective flows that can help or
hinder its progress; in general, whether the tube can rise coherently, and
if so at what rate, is a function of both the surrounding convective
velocity field and the properties of the flux tube itself.  In many simple
models, the terminal velocity attained by the tube is related to the Alfven
velocity ($v_a = B/\sqrt{4 \pi \rho}$), multiplied by factors that depend on the
size of the flux tube and the background stratification.

How such flux tubes might arise is still a matter of some uncertainty.
Some authors have argued that the natural state of the dynamo is a
collection of fibril magnetic tubes -- e.g., \citet{parker1984} argued that
this configuration is a minimum-energy state for a given field
strength. Others have shown (as demonstrated below) that even smooth layers
of magnetic field may break up and yield more compact structures that
resemble flux tubes.  Still other simulations have self-consistently
generated magnetic structures from the combined action of convection and
shear \citep[e.g.,][]{nelson_etal2011, nelson_etal2014} that rise due to
buoyancy and advection by surrounding flows, in a manner somewhat similar to simple
flux tubes.  In any case, it has proven useful in many contexts to study
the properties of such idealized tubes as a proxy for more complex field
configurations.

Parker (1975) pointed out that the rapid rise of such tubes poses major
problems for simple models of the solar dynamo.  The rise time of an
isolated tube is estimated to be much shorter than the length of the solar
cycle, and shorter than the timescale on which the field could plausibly be
regenerated by stretching amidst the convection zone.  This led Parker to
argue that the global solar dynamo might be situated at the base of the
convection zone (where rise times are larger), and to point out that fields
could be amplified to much greater strengths (before becoming susceptible
to buoyancy instabilities) in the stably stratified region below.  This,
coupled with the later discovery using helioseismology of a ``tachocline''
of shear at the interface between the convective envelope and radiative
interior \citep[e.g.,][]{thompson_etal1996}, helped lead to the
now-prevalent ``interface dynamo'' paradigm for global field generation in
the Sun \citep[see, e.g.,][]{park93, ossen2003, miesch2005_lrsp}.  This
picture has been critically reexamined by some in recent years, with
various numerical simulations suggesting that fairly strong large-scale
magnetic fields could be generated within the convection zone while
evidently avoiding severe losses by magnetic buoyancy \citep[see, e.g.,
  discussions in][]{brun_etal2013, brown_etal2011, brandenburg2005}.  But
the basic reality of magnetic buoyancy is undisputed, even though downward
``magnetic pumping'' \citep{tobias_etal2001} and other effects can limit its
effectiveness in some circumstances.

Below, we essentially adapt the argument of Parker (1975) for the case of
very strong fields in fully convective stars, incorporating what is now
known about the rise of simple flux tubes and the instability of smoother
field distributions.  We begin by calculating the rise time of the simple,
thermally isolated flux tube described above, then summarize results from
more complex calculations that incorporate the effects of radiative
transfer, interaction with the surrounding turbulence, and other effects.
We describe the conditions under which smooth field distributions are
unstable to similar buoyancy instabilities, and apply this to conclude that
some previously studied profiles for the interior magnetism in low-mass
stars are unstable.  We then estimate plausible regeneration times for the
magnetism under a variety of assumptions, and use this to assess the
strength and size of fields that could likely be regenerated faster than
they are lost from magnetic buoyancy.

\subsection{Simple estimates of rise times}

We first consider the rise of an unbent, untwisted, isolated flux tube in
thermal equilibrium with its surroundings.  It is worth noting that this is
not in itself an instability in the usual sense (there is no initially
stable configuration that is being perturbed); nonetheless this simple
configuration has many features in common with the more complex scenarios
described below. In this case, for the flux tube to remain in pressure
equilibrium with its surroundings, we must have
\begin{equation}
  P(z) = P_i(z) + B^2/8\pi
\end{equation}
with $P_i$ the pressure internal to the tube and $P$ the external
pressure.  If the temperatures inside and outside the tube are the same,
then the density within the tube must be smaller than that of the
surrounding medium, 
\begin{equation}
  \rho(z) - \rho_i (z) = \rho(z) B^2(z)/8 \pi P(z),
\end{equation}
and this reduced density results in a buoyant force per unit length $F \sim
\pi a^2 g (\rho - \rho_i)$, with $a$ the cross-sectional radius of the
tube.  Parker (1975) assumed the upward acceleration of the tube is
resisted primarily by the aerodynamic drag, $F_d = \frac{1}{2} \rho u^2 a
C_D$, with $C_D \sim 1$ the drag coefficient.  With this assumption, the
terminal velocity occurs for $F= F_D$, and hence the rise velocity is
\begin{equation}
  \label{eqn:risetime}  
  u = V_a \left(\frac{\pi a}{C_D H_p} \right)^{1/2}
\end{equation}
where $V_a = B/\sqrt{4 \pi \rho}$ is the Alfven speed and $H_p = P/ (g \rho)$ is the
local pressure scale height.  Later authors have adopted many other
prescriptions, considering for example the turbulent diffusivity of the
medium as the primary impediment to the tube's rise, but for the very
strong fields considered here these changes have only minor effects on the
total rise time.

In Figure 1, we use this simple estimate of the rate of rise of isolated
flux tubes to assess the rise time $t_{\rm rise} = (R -r)/u$ for tubes of
various sizes and strengths.  We have adopted stratifications from a 1-D
model of a 0.3 solar-mass M-dwarf, provided for us by Isabelle Baraffe
and computed following \citet{cb1997}. For simplicity and illustrative purposes we will in some discussions take $H_p$ to have
a constant, representative value equal to its value at $r = 0.25 R$ ($H_p
\approx 8 \times 10^9$ cm), but the calculations shown in Figure 1 include
the spatial variation of all quantities, integrated numerically to
calculate the rise time for tubes starting at the origin.  The rise time decreases with increasing
field strength, and is longer for small-scale fields than for large-scale
ones.  This is essentially because the rise time in this simple model is
set by a balance between drag forces (which increase with the surface area
of the tube) and buoyancy forces (which increase with its volume), so thick
tubes rise faster than thin ones in accord with their larger ratio of
volume to surface area.  In what follows, it is useful to note that the
rise time for deep-seated fields in this model is given by
\begin{equation}
  t_{\rm rise} \sim \frac{R}{u} = \frac{R}{C a^{1/2} B}
\end{equation}
where $C=1/\sqrt{4 \rho H_p C_D}$ varies slowly with radius but is
typically about $6 \times 10^{-7}$ in cgs units.

\begin{figure}
\center
\plotone{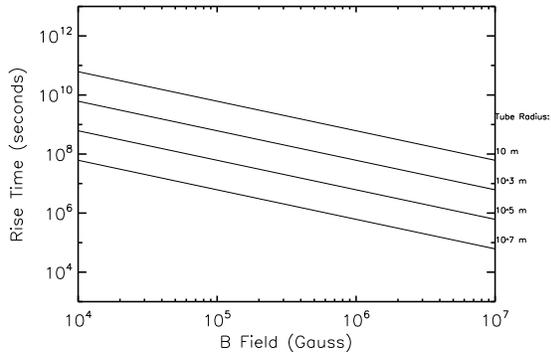}
\caption{Simple estimate of rise time (eqn \ref{eqn:risetime}) for isolated flux tubes of
  indicated sizes and strengths.  Stronger, larger-scale tubes rise more
  quickly than weak, small-scale fields. }
\end{figure}

This simple calculation neglected heat transfer between the rising tube and
the surrounding medium, among other effects.  Many other authors
\citep[e.g.,][]{schuessler1977, spruit1981, spruit_vanB1982,
  morenoinsertis1983, morenoinsertis1986, fan_1993, caligari_etal1995, caligari_1998} have carried out more sophisticated
calculations that examine the behaviour of such tubes under different
assumptions.  If, for example, the tube does not equilibrate instantly to the
surrounding temperature as it rises, but instead expands adiabatically
along its ascent, then its rate of rise depends on the properties of the
external medium.  If the medium is unstably stratified (i.e., convective),
then the tube still must rise indefinitely; if instead the surroundings
are isothermal, for instance, it is possible for the flux tube to
reach the same density as its environment and hence attain mechanical
equilibrium.  In practice, tubes are likely neither to equilibrate
instantly to their surrounding temperature nor to behave perfectly
adiabatically; the reality will be somewhere between these two extremes,
determined by the rate at which heat flows in or out of the tube.  It is
possible to show that the rise time is fastest for tubes that rise
adiabatically, and slowest for those that adjust instantly to the
surrounding temperature (see, e.g., Moreno-Insertis 1983, Fig 1).  The rise
times displayed in Figure 1 are thus arguably somewhat conservative
estimates of the rate at which isolated tubes might rise.  

Some additional insight can be gained by considering the case of a thin
magnetic tube initially in thermal or mechanical equilibrium with its
surroundings, and asking under what circumstances this equilibrium is
stable.  For mechanical equilibrium to be possible in the first place,
initially the tube must (for the reasons given above) be at a different
temperature than its surroundings, so in this case stronger fields have the
slightly bizarre effect of forcing the initial tube to be cooler.  In this
special case, it would be possible to maintain very strong fields even
amidst an unstably stratified region if the tube were required to remain
straight \citep[see][]{spruit_vanB1982, morenoinsertis1983}.  But if the
field lines are allowed to bend, then (as noted by \citealt{parker1955},
analyzed by \citealt{spruit_vanB1982} analytically, and confirmed
numerically in many later papers) a flow away from the crests of the tube
tends to enhance the buoyancy of those regions, leading again to
instability on sufficiently long wavelengths.

As a flux tube rises, it will be affected by the convective flows in a
variety of ways.  Arguably the simplest approach, as adopted above, is to
assume that the flux tube's motion is resisted by aerodynamic drag.  Other
authors have modeled the braking effect of the surrounding turbulence in
other ways, for example as an eddy viscosity
\citep[e.g.,][]{unno_ribes1976, schuessler1977, morenoinsertis1983},
leading in some cases to significantly longer rise times.  The drag in the
turbulent viscosity models tends to be greater than that in models assuming
aerodynamic drag if the velocity of the rising flux tube is small compared
to the convective velocity (e.g., Moreno-Insertis 1983); for fields (like
those considered here) that are much stronger than the convective flows,
turbulent viscosity cannot greatly impede the flux tube's rise.  Later work
using numerical simulations has extensively investigated the rise of both
tubes and more general magnetic structures along with (in some cases) their
interaction with convective flows, whether imposed, self-consistently
generated, or simply parametrized.  Examples are provided by
\citet{hughes_falle1998, wissink_etal2000, cattaneo_etal2006, cheung_etal2006, jouve_brun2009, favier_etal2012, barker_etal2012,
  pinto_brun2013, martinezsykora_etal2015pre}. The
processes by which magnetic fields are more generally expelled from regions
of active convection has also been extensively studied; see
\citet{tobias_etal1998, tobias_etal2001, tobias_etal2008, weiss_etal2004} for examples.
In general, for flux tubes to survive passage through the convection
zone in these models they must exceed the value in equipartition with the convection by a
geometrical factor of order $H_{\rho}/a$, where $a$ is the cross-sectional
radius of the flux tube: in other words, at fixed field strength, small
flux tubes are more easily held down by the convection.

The qualitative trends revealed by comparatively simple estimates -- that
strong, thick tubes rise faster than weak or thin ones -- largely appear to
be realized for somewhat more complex flux tube configurations as well, even when a
panoply of effects not present in Parker's original formulation are
included.  Further, while the thin flux tube approximation itself suffers from significant limitations \citep[see, e.g., discussion in][]{hughes2007}, its estimates appear to capture the overall rate of rise of buoyant fields reasonably well, at least when compared to full MHD simulations of rising field structures in certain regimes \citep{cheung_etal2006}. We therefore adopt these estimates  in much of 
our discussion here, while recognizing that variations in how the tube
exchanges heat with its surroundings, interacts with the convection, and
expands with height, are all likely to affect our conclusions
quantitatively at some level.  In \S 3, we will turn to our own thin
flux-tube MHD simulations, which include some of these effects (albeit
still in a very simplified way)  to assess how robust our estimates may be.

\subsection{Instability of magnetic layers}

Just as isolated flux tubes represent one limit of possible field
configurations, another widely studied limit involves a global distribution
of magnetism, varying initially with depth in some specified way.  Here we
briefly review the extensive literature on buoyancy instabilities of both
discontinuous layers of field (i.e., ``slabs'' of magnetism) and continuous
field distributions. 

In general, an atmosphere that contains a layer of magnetic field
underlying an unmagnetized (or less-magnetized) fluid is unstable to
buoyancy instabilities.  In essence, the presence of the magnetic field
makes the fluid ``top-heavy,'' because heavier (less-magnetic) gas is being
supported by lighter (more magnetic) gas, so these instabilities often take on the same
character as classical Rayleigh-Taylor instability.

It is useful to distinguish between two broad types of disturbances, known
as ``undular'' and ``interchange'' modes.  Interchange modes are those that
involve no bending of field field lines (i.e., $k$ is perpendicular to
$B$), while purely undular modes have $k$ parallel to $B$.  In general,
perturbations may be fully 3-D and have elements of both undular and
interchange modes.  The growth rates of the interchange modes increase
uniformly with increasing wavenumber, limited ultimately only by
dissipation.  The undular modes, on the other hand, have maximum growth
rate at finite and relatively small wavenumbers (i.e., large spatial
scales), because small scales are stabilized by magnetic tension.  Because
of this, the interchange modes generally have much faster linear growth
rates than the undular modes \citep[e.g.,][]{nozawa2005, cheung_isobe2014}.
The nonlinear stage of the instability, however, may be dominated by the
undular (or mixed) modes in many circumstances.  The linear growth rates
for these types of instabilities are of order $H_p/v_a$ in the absence of
diffusion and rotation \citep{acheson1979}, implying that for the field
strengths under consideration here, they would develop on timescales that
are short relative to both the large-scale convective overturning time and to the rise
time of small flux tubes.

For the interchange modes, it is straightforward to show that a necessary
and sufficient condition for instability is
\begin{equation}
  \label{eqn:interchange}
  \frac{v_a^2}{\gamma H_{P}}\frac{d}{dz} \ln \left(\frac{B}{\rho}
  \right) < N^2
\end{equation}
where $N$ is the Brunt-Vaisala (buoyancy) frequency and other symbols take
their usual meanings \citep{acheson1979, hughes_proctor1988, newcomb1961}.
Thus, even in a convectively \emph{stable} atmosphere, instability will
occur if $B/\rho$ decreases rapidly enough with height; in a neutrally or
unstably stratified layer, \emph{any} decrease in $B/\rho$ with height is
unstable.

The window of instability for undular modes is even broader.  These modes
(in which the field lines are bent rather than remaining straight) can
occur in circumstances where the interchange modes are still stable,
essentially because in the undular case the instability can (by compressive
motions along the field lines) avoid doing extra work against magnetic
pressure while tapping into the potential energy of the stratification
\citep{hughes_cattaneo1987}.  These modes are thus the smooth-field analogue to
the ``bent flux tubes'' discussed above, and (as in that case) are unstable
in some regimes where the interchange modes are not. As shown by
\citet{newcomb1961} using an energy
priciple, and analyzed by \citet{parker1966} using normal modes (see also \citealt{thomas_nye1975}), instability occurs (in the absence of dissipation) if
\begin{equation}
  \label{eqn:undular}
  \frac{1}{\gamma H_{P}}\frac{d}{dz} \ln B
   < \frac{k_y^2 k^2}{k_x^2} + \frac{N^2}{V_a^2}
\end{equation}
where $k_x$ and $k_y$ are horizontal wavenumbers (with $k_y$ the wavenumber
along the tube; recall that for the interchange modes $k_y=0$), and $k$ is
the total wavenumber.  Thus while the interchange modes require $B/\rho$ to
decrease with height (or increase with depth), the undular modes require
only that $B$ decrease with height.  On the other hand, they are also
stabilized at short wavelengths by magnetic tension; for field
distributions that vary slowly with height, only global-scale perturbations
are likely to be unstable.

These instabilities have, like their isolated flux tube equivalents, been
studied extensively using numerical simulations.  The linear growth and
subsequent evolution of pure interchange modes (sometimes called the
``magnetic Rayleigh-Taylor'' instability) were examined by
\citet{cattaneo_hughes1988}, for example; the behavior of the pure undular mode (Parker instability) was
studied by \citet{shibata_etal1989, shibata_etal1989b} and \citet{fan2001},
among others.  Many subsequent papers have
examined the general nonlinear behavior of layers of field subject to
arbitrary (mixed-mode) perturbations \citep[e.g.,][]{matsumoto_shibata1992,
  matthews_etal1995, wissink_etal2000, kersale_etal2007}.  In
general, the effect of these instabilities is to cause an initially uniform
layer to break up into smaller features that then evolve nonlinearly under
the competing influences of buoyancy, magnetic tension, and other effects.
Thus for example in the solar dynamo, it is widely thought that these processes may play a role in
generating magnetic structures akin to ``flux tubes'' from an initially
smooth field in the tachocline \citep[e.g.,][]{cheung_isobe2014}.  

It is helpful to consider the magnetic field profiles studied in FC2014 in
light of these instability criteria.  For specificity, consider their
``dipole'' field profile, which reaches a specified maximum amplitude (of
order 10$^7$ G in some cases) at a radius $r_{t}=0.15 R$, and falls off
like $1/r^{3}$ away from that point.  By inspection, this profile is
unstable to undular modes (which require only that $B$ decrease with
height) at all radii exterior to $r_t$, but these modes could in principle be suppressed
at short wavelengths.  In Figure $2$, we further compare the gradients of
the magnetic field and of density, $d (\ln B)/ dr = B^{-1} (dB/dr)$ and $d (\ln
\rho)/dr$; recall that if the field declines more rapidly than the density
(i.e., $d (\ln B) /dr$ is greater in magnitude than $d (\ln \rho) /dr$),
the profile is unstable to pure interchange modes as well.  For
convenience, we have expressed these as scale heights, plotting $H_B =
-B/(dB/dr)$ and $H_{\rho} = -\rho/(d\rho/dr)$; instability to interchange
modes thus requires $H_B < H_{\rho}$.  In the region
exterior to $r_t$ (i.e., where the field is falling off with height), $H_B
= r/3$, whereas density falls off less rapidly out to a fractional
radius $r \approx 0.6 R$.  Thus the profile is unstable to interchange
modes (at all wavelengths) over a substantial fraction of the interior.
The same would be true of the ``Gaussian'' profiles also employed in
FC2014, for which $B$ declines more steeply than $\rho$ throughout
virtually the entire domain exterior to $r_t$.  We therefore conclude that
a smooth field distribution whose radial variation is given by either of
these profiles would be unstable to both interchange and undular modes of
magnetic buoyancy instability over a large fraction of the interior, and
would therefore break up into smaller-scale magnetic structures.  Although
the nonlinear evolution of these structures is likely to be complex, prior
3-D simulations of these instabilities have suggested that the ensuing
behavior resembles that of simple flux tubes in some important respects,
namely that the strongest-field regions tend to rise buoyantly unless they
are ``pumped'' downwards by convection.

\begin{figure}
  \epsscale{1.2}
\center
\plotone{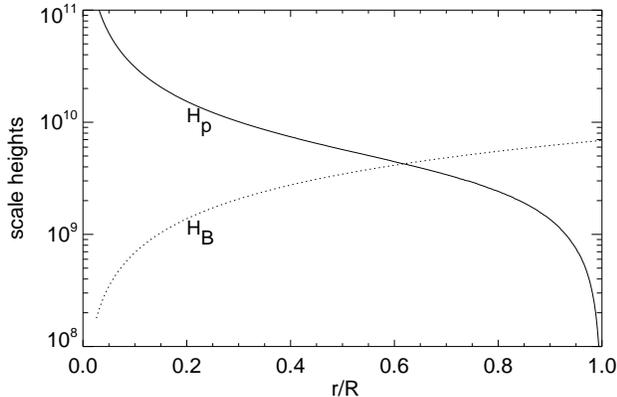}
\caption{Analysis of the instability of previously proposed smooth magnetic field profiles  via undular and interchange
  modes. Shown is a comparison of $H_{\rho} = -\rho/(d\rho/dr)$ in a 1-D
  stellar model and $H_{B} = -B/(dB/dr)$ for part of the FC2014 ``dipole''
  field profile.  (The FC2014 profile departs from this curve for $r/R <
  0.15$, where $B$ increases with $r$.) The regions where $H_{B}
  < H_{\rho}$ (i.e., interior to $r \approx 0.6 R$) are potentially unstable to both undular and interchange
  modes.  All regions exterior to $r/R = 0.15$ (i.e., the radius at which
  the field strength begins to decline with radius) are unstable to undular modes at some
  wavelengths.  }
\end{figure}

Because the specific radial profiles employed in FC2014 are somewhat
arbitrary, it is also instructive to examine the possible instability of
more general field distributions.  Any profile $B(r)$ that increases with
depth will be unstable to undular modes, but the linear growth rates of
these modes can be relatively slow (since they occur only on large spatial
scales), and we find it difficult to predict with certainty whether such instability would
act ultimately as a loss mechanism for magnetic energy in the presence of
many other competing effects.  Consider, then, the instability to
interchange modes, which occur on all scales and hence grow more
rapidly.  If we suppose that $B(r)$ increases smoothly from an observed
surface value $B_{\rm surf}$ to some interior maximum $B_{\rm max}$,
instability to interchange modes is inevitable if $B_{\rm max}/B_{\rm
  surf}$ exceeds a value of order $\rho_{\rm max}/\rho_{\rm surf}$.  The
radial surface on which we can constrain $B_{\rm surf}$ is not just the
stellar photosphere (where $\rho$ is very small), but somewhat deeper.  In
the Sun, for example, although the radial extent of large active regions is
still a matter of considerable debate, sunspots are coherent enough to
support magneto-acoustic modes of oscillation down to a depth of at least
10 Mm \citep[e.g.,][]{kosovichev2002, kosovichev_duvall2006}; this in effect constrains the
field in the outer 10 Mm of the solar convection zone to be no stronger
than the fields observed in spots. In convective M-dwarfs, observations
suggest surface fields of no more than a few kG
\citep[e.g.,][]{johnskrull_valenti1996, donati_etal2006}, and it is reasonable to assume that no stronger fields than
this exist in the outer few Mm of the star.  For specificity, if the field
at 0.95$R$ (i.e., less than 10 Mm below the surface) is constrained to be
no more than 5 kG by observations, then $B(r)/\rho(r)$ cannot exceed its
value at that radius without triggering instability to interchange modes.
In a sample 1-D model of a 0.3 solar-mass main-sequence M-dwarf provided to us by
Isabelle Baraffe \citep{cb1997}, the density at 0.95$R$ is about 0.9 g cm$^{-3}$, whereas in the interior it
reaches a maximum of about 100 times this ($\rho_{\rm max} = 92$ g
cm$^{-3}$), so to avoid instability $B$ must not exceed values of order
$100 \times B_{\rm surf} \approx 5 \times 10^5$ G anywhere in the
interior.  If this limit is violated, it is difficult to avoid the
conclusion that even smooth layers of field would quickly break up from buoyancy
instability, with a subsequent nonlinear evolution that includes buoyant
rise towards the surface.

\subsection{Regeneration times for the magnetism}

The buoyant rise of strong magnetic fields represents a mechanism by which
these fields may be ``lost'' from the deep interior.  But we also expect
that convection, rotation, and shear within the star will together act to
regenerate the field through dynamo action.  In this section, we briefly
examine the timescales on which this regeneration might occur.

A simple estimate of the characteristic timescale for field growth can
be derived by examination of the induction equation, which in the absence
of dissipation is
\begin{equation}
  \frac{\partial{\bf B}}{\partial t} = \nabla \times ({\bf v} \times {\bf B}) = - {\bf v}
  \cdot \nabla {\bf B} + {\bf B} \cdot \nabla {\bf v}
\end{equation}
where all symbols take their usual meanings.  
Qualitatively, the first term on the right hand side represents advection of
the magnetism by the flow; the second term captures stretching of
fieldlines by a non-uniform flow.

If we first suppose that the field and flow can each be characterized by
just one spatial scale ($L$) and one timescale $t_{\rm gen}$, then the
induction equation suggests
\begin{equation}
  \frac{B}{t_{\rm gen}} \sim \frac{V B}{L} \rightarrow t_{\rm gen} \sim
  \frac{L}{V}.
\end{equation}
Thus in the absence of dissipation and any nonlinear feedback on the flow,
we expect that the field will grow on a timescale of order the convective
overturning time.  Much more sophisticated limits can be derived by turning
to the theory of fast dynamos -- i.e., dynamos whose growth rate is
non-zero in the limit of zero diffusion \citep[see, e.g.][for
  review]{childress_gilbert1995}.  It was conjectured by
\citet{finn_ott1988}, and later proven by \citet{klapper_young1995} in
certain circumstances, that the growth rate for kinematic dynamos is
bounded above by the topological entropy of the flow (which is in turn a
bound on the Lyapunov exponents of the flow, and physically is related to
the rate at which trajectories in a chaotic flow diverge).  In practice,
applying this limit is not particularly useful for our purposes: the
problem we consider here is neither purely kinematic nor free of diffusion,
and moreover there is no obvious way to estimate the topological entropy of
an arbitrary convective flow.  Fortunately, numerical simulations of dynamo
action in a wide variety of contexts suggest that the simpler estimate
above ($t_{\rm gen} \sim t_{\rm dyn}$, suitably defined) is a reasonable
upper bound on the rate at which magnetic energy grows in the kinematic phase
\citep[e.g.,][]{cattaneo_hughes2006_jfm}.  Field growth in the nonlinearly saturated phase is likely to be considerably slower.

We will adopt this simple bound in our discussion below, but a few
additional points require specification.  These relate to what is assumed
about the flow, and about the lengthscales on which it and the field are
assumed to vary.

We will assume, first, that the growth of fields is limited by appropriate
{\sl convective} timescales, rather than by timescales associated (say)
with some unknown internal shear flow.  This is not because shear is
guaranteed to be dynamically unimportant; indeed, as we note below, it is
conceivable that internal differential rotation might play as significant a
role as the convection.  But no purely toroidal flow (such as that
represented by differential rotation) can act as a dynamo on its own (see
discussions in \citealt{bullard_gellman1954}, \citealt{jones2008}).  Another flow must act
to generate poloidal field from toroidal, and in general the overall rate
of growth will be limited by the slower of these.  The most plausible
candidate for the latter step is the convection, so we assume it is the
stretching properties of this flow that will limit how quickly a general
field can be rebuilt.

Convection in stellar interiors occurs over a wide range of spatial and
temporal scales, so estimates of the convective overturning time (and hence
the field regeneration time) are fraught with some uncertainty.  The
simplest model is that the regeneration process must ultimately occur on
the turnover time of the comparatively large-scale, slowly overturning
eddies in the convection, since these carry most of the energy.  Mixing
length estimates \citep[e.g.][]{hansen_kawaler1994, thompson_duncan1993}
and simulations of stellar/planetary convection
\citep[e.g.,][]{abbett_etal1997, browning_etal2004, meakin_arnett2007, viallet_etal2013}
both suggest that these eddies should have velocities that scale roughly as
$v_c \sim (F/\rho)^{1/3}$, with $F$ the energy flux that must be carried by
convection, and should have typical sizes of order a pressure scale height
(which is in turn of order the stellar radius $R$) in the deep interiors of
these stars. Written explicitly, we have
\begin{equation}
  t_{\rm gen}^{\rm slow} \sim \frac{d}{v_c} \sim \frac{d}{\left(F/\rho
    \right)^{1/3}}, 
\end{equation}
defining the depth $d = R - r$.

  This estimate (which we refer to below as the ``slow'' model) involves 
  velocities in the interior of a 0.3 solar mass star that are of order a
  few m s$^{-1}$, and hence implies overturning times of order $10^{8}$ seconds, or
  about 3 years.  Turning again to Figure 1, we see that if this is the
  timescale on which fields are regenerated, very strong flux tubes (of
  more than say $10^6$ G strength) have rise times shorter than the
  regeneration time unless they have very small radii ($a \le 10^3$ m).

But the convection likely contains eddies with a wide range of spatial
scales, and the smallest of these eddies plausibly overturn much more
quickly than the large-scale motions modeled in MLT.  Although these
smaller-scale eddies do not contain much energy, it is worth considering
the timescales associated with field regeneration on these smaller scales.
This is also the timescale on which a large-scale flow acting on a
pre-existing {\sl small-scale} field could produce fluctuations on that
same small scale.  We consider a simple model where a large-scale velocity
field $v_c$ is taken to act on fields at all scales $a$, with a
characteristic regeneration time
\begin{equation}
  t_{\rm gen}^{\rm fast} \sim \frac{a}{v_c} \sim \frac{a}{\left(F/\rho \right)^{1/3}}.
\end{equation}
We retain this estimate mainly as an upper limit to how quickly fields could
be regenerated for a short time by large-scale flows (e.g., shear,
discussed more below).  We refer to this estimate as the ``fast
eddy/shear'' model.

It is also possible to define a scale-by-scale overturning time taking into
account the variation of the flow with scale; that is, assuming that fields at
each scale are built by motions at the same scale.  The scale dependence of the
velocity in rotating, magnetized convection is still a matter of
considerable debate \citep[e.g.,][]{kraichnan1965, grappin_etal1982, goldreich_sridhar1994,
  goldreich_sridhar1995, biskamp_muller2000, boldyrev2006,
  beresnyak_lazarian2010, perez_etal2014}, and in all cases these small-scale eddies presumably
possess less kinetic energy than somewhat larger-scale
flows (since the velocity $v^l$ on scale $l$ decreases with
decreasing $l$), so our estimates here are highly uncertain.  We consider a
scale-dependent velocity field $v^l$ that scales with $l$ as
\begin{equation}
  v^l = v^0 \left(\frac{l}{l^0} \right)^{\alpha}
\end{equation}
where $v^0$, $l^0$ refer to the velocities on some large length scale
$l^0$.  The power-law index $\alpha$ would be $1/3$ for turbulence obeying a
Kolmogorov-like cascade.  In this model the
turnover time, which we will take as representative of the regeneration
time for fields on scale $a$, scales like
\begin{equation}
  t_{\rm gen}^{\rm } \sim \frac{a}{v_c^a} \sim \frac{a}{\left(F/\rho
    \right)^{1/3} (a/R)^{\alpha}},
\end{equation}
where we have further assumed that the large-scale eddies (whose
scale-dependent amplitude $v_c^a$
is given by mixing-length theory) have a spatial scale of order the stellar
radius $R$ (which is of order the pressure scale height in the deep
interior).  We refer to this as the ``cascade'' model below.

\subsection{Resulting constraints on field strengths and morphologies}

Our analysis thus far has suggested that fields of the strength envisioned
in some previous models (MM01, FC14) are likely to be unstable.  If the
magnetism consists of smooth layers, it is apt to break up into
smaller-scale structures unless the stability criteria outlined above are
satisfied; these instabilities are generally very rapid \citep[e.g.,][]{acheson1979}. If the field
consists of bundles of discrete flux tubes it is again unstable, on a
timescale that varies with the geometry and strength of the field.

In both cases, the instability acts as a loss mechanism for fields in the
deep interior, acting over some characteristic time $\tau_B$.  If the field
is not regenerated on a comparable timescale, then the overall level of
magnetic energy must decline.  For the proposed magnetic field profiles to
represent steady-state solutions at the present day, the star would need to
produce field in the interior at least as quickly as it is carried away by
buoyancy instabilities.

Denoting the buoyant rise time for fields of given strength $B$ and spatial
scale $a$ by $t_{\rm rise}$, and the regeneration time for the same fields
by $t_{\rm gen}$, we can therefore equate $t_{\rm rise} \sim t_{\rm
  gen}$ to find a crude estimate of the maximum field that could be
sustained indefinitely.  In Figure 3, we show this maximum field strength
as a function of radius for simple flux tubes (i.e., tubes obeying eqn \ref{eqn:risetime}) of various sizes, assuming that the regeneration timescale is given by
the overturning time of the slow, large-scale convective eddies (i.e., the
``slow eddy'' model $t_{\rm gen}
\sim d/v_c$, with $v_c$ varying with depth as described above).  In this
case
\begin{equation}
  \frac{t_{\rm gen}}{t_{\rm rise}} \sim \frac{d/v_c}{d/Ca^{1/2} B} = \frac{Ca^{1/2} B}{v_c} \sim 1
\end{equation}
where $C$ is as described earlier, so in equilibrium the field in tubes of size $a$ is given by
\begin{equation}
  B \sim \frac{v_c (r)}{C(r) a^{1/2}}.
\end{equation}
This is, to order of magnitude, the same as the {\sl minimum} field needed
for fields at a given scale to overpower the effects of downward magnetic
pumping by the convection.  For example, \citet{fan2009} estimates the latter as
\begin{equation}
B \ge \frac{2 C_D}{\pi} \left(\frac{H_p}{a} \right)^{1/2} B_{\rm eq}
\end{equation}
where $B_{\rm eq} \approx \sqrt{4 \pi \rho} v_c $ is the field strength in equipartition with large-scale
convective eddies. In essence, this
just says that fields that are buoyant enough to overcome downward pumping by
convection will always rise faster than they can be regenerated by the
large-scale eddies.  In this limit, to maintain fields of order $10^7$ G
would require that the field be composed of tubes with scale $a \le 100$
m; conversely, for fields of scale $a=10^7$ m, only fields with strengths
less than about $10^5$ G could be maintained.  These estimates vary
slightly depending on what is assumed about the radial variation of $H_p$
(and with it the quantity $C$) and the turnover time; if we ignore the
radial variation of $H_p$, and adopt a constant value representative of the
deep interior instead, the maximum allowable field is reduced by a factor of about
three.  

\begin{figure}
  \epsscale{1.2}
\center
\plotone{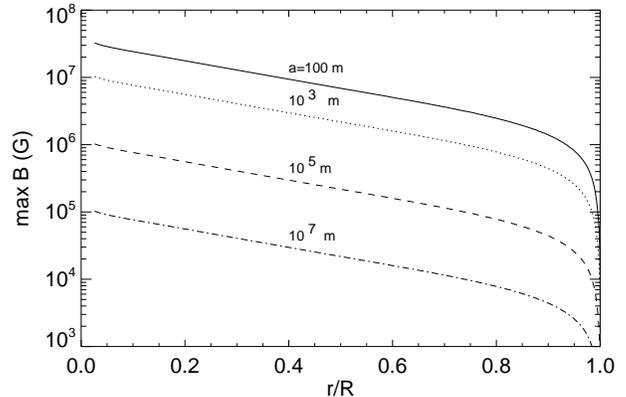}
\caption{Maximum field strength achievable if large-scale eddies with
  turnover time $\tau \sim d/v_c$ must regenerate field lost due to
  magnetic buoyancy during the same time. Limits are shown as function of
  radius for varying flux tube sizes $a$, using the simplest rise
  time estimate (eqn 3). The maximum sustainable field is larger if the
  field scale $a$ is small, because the rise time for those fields is
  long. }
\end{figure}

If instead we make the generous assumption that scale-by-scale regeneration
of the field occurs on a timescale $t_{\rm gen} \sim a / v_{c}$, taking
$v_c$ to be a constant (what we called the ``fast eddy/shear'' model above),
then much stronger fields can be maintained on small scales.  In this case
\begin{equation}
  B \le \frac{v_c d}{C a^{3/2}},
\end{equation}
so the maximum field is faster than in the ``slow'' estimate by a factor
of $d/a$, which is large for small $a$.  In practical terms this
means that fields with strength $10^7$ G are too buoyant unless $a \leq
10^6$ m.

The maximum field strength in the ``cascade'' model, which has velocities
$v_c^a(r)$that fall off with decreasing spatial scale according to a power law, is
intermediate between these two limits at given $a$ for reasonable choices of the
power-law dependence.  Specifically, the choice $\alpha=1$ (which gives a
constant scale-dependent overturning time $\sim a/v^a$) is akin to the
``slow eddy'' model, whereas the choice $\alpha=0$ (implying the local
overturning time decreases linearly with $a$) corresponds to the ``fast
eddy/shear'' case.  A Kolmogorov-like cascade corresponds to $\alpha=1/3$,
and the Iroshnikov-Kraichnan spectrum would correspond to $\alpha=1/4$, in both
cases yielding field estimates that fall between the ``slow'' and ``fast''
cases shown in these figures. Written explicitly, the maximum field is of
order
\begin{equation}
  B \le \frac{v_c^R}{C a^{1/2}} \left( \frac{d}{a} \right)
  \left(\frac{a}{R} \right)^{\alpha},
\end{equation}
i.e., it differs from the ``slow eddy'' estimate by the factors $(d/a)
(a/R)^{\alpha}$. (The ``slow eddy'' expression quoted earlier is slightly different from
that derived by taking $\alpha=1$ here, simply because we have assumed
the relevant scale in the turbulent cascade is $(a/R)$ rather than $(a/d)$.)  

Equivalently, we can also use these estimates to construct limits on the
maximum characteristic spatial scale of the field, if it is to be stable
against buoyancy or regenerated faster than it buoyantly rises.  In general, for the ``cascade''
model of field regeneration with a scale-dependent overturning time $t \sim
l/v_c(l)$ and taking $v_c(l) \propto l^{\alpha}$, we have
\begin{equation}
  a_{\rm max} = \left[\frac{d}{R^{\alpha}} \frac{v_c^R}{C B}
    \right]^{\frac{2}{3 - 2 \alpha}}
\end{equation}
where as before $C=1/\sqrt{4 \rho H_p C_D}$ and we are denoting $v_c^R(r)$ as
the velocity on spatial scale $R$ at location $r$.  It is perhaps more instructive to
rewrite this assuming that $H_p \approx R$ (true to order of magnitude in the deep interior of a
low-mass star), $C_D \approx 1$, $d \approx R$, and denoting $B_{\rm eq} \sim \sqrt{4 \rho}
v_c(R)$ as the magnetic field strength in equipartition with the
largest-scale convection, so that
\begin{equation}
  \label{eqn:amaxequipart}
  a_{\rm max} \sim R \left(\frac{B_{\rm eq}}{B} \right)^{\frac{2}{3 - 2 \alpha}}.
\end{equation}

Written in this way, it is clear that fields of order the equipartition
strength are stable against buoyancy, or rise more slowly than they are
regenerated, on all spatial scales.  Stronger fields (relative to
equipartition) are more subject to rapid buoyant losses, but these can be
avoided if the characteristic size scale of the field is small enough.  How
small is small enough depends on the assumed characteristics of the
regeneration process.  Here $\alpha = 0$ corresponds to the extreme ``fast
eddy/shear'' model in which the scale-by-scale turnover time decreases
linearly with decreasing spatial scale; in this case $a_{\rm max} \propto
(B_{\rm eq}/B)^{2/3}$.  If regeneration on small scales is somewhat slower,
as for a Kolmogorov-like cascade ($\alpha \sim 1/3$) or the ``slow eddy''
models considered above ($\alpha = 1$), then $a_{\rm max} \propto (B_{\rm
  eq}/B)^{6/7}$ or $(B_{\rm eq}/B)^{2}$ respectively.

\begin{figure}
  \epsscale{1.2}
\center
\plotone{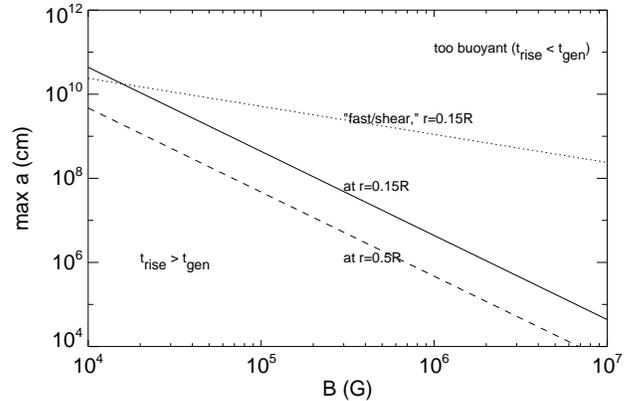}
\caption{Maximum size of flux tubes at any given field strength for which
  the rise time is longer than the generation time, for the ``fast/shear''
  and ``slow'' models considered above.  Only very
  small-scale fields are permissible at high field strengths, because
  large-scale fields rise more rapidly than they can plausibly be
  regenerated. The solid and dashed lines also correspond to the
  largest-scale fields that could be pumped downward by eddies at a given
  field strength at that depth (see text).  The limits vary somewhat with
  radius; limits at two representative depths are illustrated for the
  ``slow'' regeneration model. \label{fig:amax}}
\end{figure}

These
limits are assessed in Figure \ref{fig:amax}, which shows the maximum size $a$ for which
thin flux tubes have $t_{\rm rise}$ longer than $t_{\rm gen}$.  These
limits vary somewhat with radius, so we have chosen two representative
spots in the interior (at $r=0.15R$, where the FC2014 ``dipole'' field
reaches its maximum, and $r=0.5R$) for some estimates. The solid
and dashed lines give the limits when the relevant overturning time is the large-scale
one (i.e., the ``slow/eddy'' model), and so also correspond to an approximate limit on fields that are ``pumped''
downward (i.e., fields above the solid line are too large in scale to be pumped
downward at any given field strength).  This corresponds to the line
\begin{equation}
  \label{eqn:amaxslow}
  a_{\rm max} = \left(\frac{v_c}{B}\frac{1}{C} \right)^2,
\end{equation}
where $C$ is as before.  The dotted line (labeled ``fast/shear'') corresponds to the more optimistic
(and in our opinion unrealistic) estimate in which the field is taken to be
regenerated on the rapid, kinematic small-scale overturning time $a/v_c$.  Here the
allowable field can be considerably larger in scale in most cases.  (An
exception occurs at very large $a > d$, because in that case the ``slow
eddy'' model actually gives slightly faster regeneration times, $t_{\rm gen}
=d/v_c$, than the ``fast'' model with $t_{\rm gen} = a/v_c$.)

We will adopt the more stringent estimate of equation \ref{eqn:amaxslow},
which we think best reflects limits on the field growth in this
regime -- and also, importantly, represents limits on a field that is not
actively maintained by dynamo action, but is instead passively responding
to buoyancy and magnetic pumping.  We note that, as recognized in FC2014, the
fields permissible at very high field strengths can only be on very small
scales.  For fields of $10^{7}$ G, still somewhat smaller than the
strongest fields considered in FC2014, only flux tubes with $a \le 2 \times
10^4$ cm are allowed; for $B=10^6 G$, $a_{\rm max} \sim 2 \times 10^6$ cm.
It is also reassuring to see that large-scale fields (with sizes of order
the stellar radii) are permissible at field strengths $10^4$ G, since
fields of roughly this size and strength have been found in numerical
simulations of low-mass stellar dynamos \citep[e.g.,][]{browning2008}.  It is
important to note that in any realistic dynamo, many scales will likely be
present; in this case, these estimates constrain the maximum {\sl average}
size of the field, but do not preclude some fields on both larger and
smaller spatial scales.

\subsection{Complicating factors: stable cores, rotation, shear}

We briefly note here a few factors that would influence our conclusions in
this section to some degree.  We consider the possible presence of a stably
stratified core, the existence of strong internal shear, and the influence
of rotation on buoyant rise times.

We have so far considered only stars that are convective throughout their
interiors; we will largely defer a detailed examination of how these
constraints scale in stars possessing radiative cores to future work.  But
it is worth commenting on a few qualitative features that arise when a
stably stratified region is present, whether that region is the result of
normal stellar evolution (as in somewhat more massive stars) or instead
arises from the stabilizing action of extraordinarily strong magnetic
fields (as envisioned by \citealt{mullan2001}).  The presence of a
small, stably stratified core slows but generally does not entirely stop
the action of the buoyancy instabilities outlined above.  There is now the
possibility of stable fields, as examined briefly in \citet{mullan2001};
however, these must still satisfy the criteria outlined in equation
\ref{eqn:interchange} or equation \ref{eqn:undular}.  Whether strong fields
are stabilizing or de-stabilizing depends on the mechanical and thermal
properties of the flux tube, but it is generally not possible to avoid
instability forever at all wavelengths.  The rise times through stably
stratified layers in general tend to be much longer than through unstably
stratified ones, because they are partly controlled by the slow rate of
radiative diffusion into the flux tube \citep[see,
  e.g.,][]{macgregor_cassinelli2003}.  But if the stably stratified region
is small and only weakly stratified, as it would likely be in the MM01
``magnetic stabilization'' scenario, then a number of other effects might
well be just as significant as magnetic buoyancy in bringing fields towards
the surface.  For example, overshooting from the overlying convection into
the putative core, and consequent entrainment, would act to mix fields
between the two regions; meridional circulations and magnetic diffusion,
though much slower, would also tend to link the two zones.  Other
instabilities not considered here (e.g., the Tayler instability, long
studied in the context of massive stars -- see, e.g., \citealt{spruit2002,
  braithwaite2006, zahn_brun_mathis2007}) would also come into play at some
level.  Some of these mechanisms are quite slow compared to buoyancy.  But
in contrast to the case where the fields are taken to be built
self-consistently amidst the convection, magnetism of this strength in a
radiative layer is usually expected only to decay: so even if the timescale
for instability is longer, its end effect as a loss mechanism for the field
is still likely to be severe.  We defer discussion of any possible
Tayler-Spruit dynamo action \citep{spruit2002} in such regions, and
likewise of dynamos driven by buoyancy and shear alone \citep{cline_etal2003} to later work.

The presence of strong internal shear could also influence the generation
of fields, but would not greatly modify our conclusions here.  It is
plausible that shear might play as great a role as convection in building
the fields, and lead to more rapid amplifications of toroidal field than
would otherwise be possible.  Indeed, although estimates of internal shear
in M-dwarfs are highly uncertain, even very small fractions of the overall
rotational kinetic energy could, if converted into differential rotation,
act more effectively than the convection to build toroidal fields.  For
example, if an M-dwarf rotating at 10 km s$^{-1}$ managed to sustain
internal shear with $\Delta \Omega/\Omega \sim 10^{-3}$, this shear would
possess an energy density comparable to (in fact somewhat greater than)
that in the convection.  Although we would generally expect internal shear
to be small in the cases investigated here, owing in part to the effects of
the strong Maxwell stresses associated with the magnetism, such
proportionally tiny amounts of shear cannot reliably be ruled out.
(\citealt{macdonald_mullan2015} have investigated a more extreme version of this
scenario, considering equipartition of the magnetism with the overall
rotational kinetic energy as one limit.  It would, however, be difficult to
reconcile such equipartition with the simultaneous requirement that the
overall angular momentum of the system be conserved.)  But rapid
amplification by strong internal shear flow is not a panacea: as noted
above, no purely toroidal flow can act as a dynamo \citep{bullard_gellman1954}.  So although energetically the differential rotation might allow for
the build up of strong fields, the timescale on which this can occur must
still be limited by the conversion from toroidal back to poloidal field,
which is likely still to be accomplished by processes no faster than the
convective overturning time.  Thus for the dynamo to reach a steady state,
the convection still must act to rebuild the fields more rapidly than
buoyancy acts to remove them.

Rotation also slows and modifies, but typically cannot stop, the rise of
buoyant magnetic flux. The influence of rotation on the buoyancy
instability of a continuous field distribution has been studied extensively
\citep[e.g.,][]{gilman1970, acheson_gibbons1978, acheson1979,
  schmitt_rosner1983, hughes1985}, as has the influence of shear
\citep[e.g.,][]{tobias_hughes2004, vasil_brummell2009}.  Rotation tends to be somewhat stabilizing, in the sense
that though in most cases it does not dramatically modify the onset of
instability of a plane layer of field, it does reduce the growth rates of
the instability \citep[e.g.,][]{acheson1979}; however, these effects are most
significant only when the Alfven speed is much less than $\sim \Omega H$,
and still generally allow instability to non-axisymmetric modes even when
rotation is quite rapid.  Even adopting a growth rate for such
instabilities typical of the rotationally-stabilized problem ($\sim
v_a^2/(\Omega H^2)$, with $H$ the scale height and $\Omega$ the frame
rotation rate) would, for the field regime examined here, typically result in
growth times that remain comfortably shorter than the other relevant
timescales in the problem (specifically, both the convective overturning
time and the buoyant rise time for small-scale fields).  The impact of
rotation on rising flux tubes has likewise been studied by many authors
\citep[e.g.,][]{choud_1987, choudhuri1989,dsilva_choudhuri1993}, with particular focus on the role that Coriolis forces play in
setting the emergence latitudes and tilt angles of emerging flux
ropes. These latitudes, and the systematic tilt of active regions observed
in the Sun (Joy's law), are thought to constitute major observational
constraints on the operation of the solar dynamo \citep[e.g.,][]{ossen2003}. Qualitatively, if
Coriolis forces are strong relative to buoyancy, rising flux tubes tend to
move parallel to the rotation axis; if they are weak, they do not greatly
alter the emergence latitudes of the tubes or produce systematic tilt
angles akin to Joy's law; intermediate between these two regimes, the
rising tubes are not deflected too poleward but still have tilt angles
similar to those observed in the Sun. For the extremely strong fields that
are the primary focus of this paper, we do not expect Coriolis forces to
radically alter the buoyant rise time or the resulting limits on field
strength and morphology.  In \S 3, we provide some more quantitative
assessments of the likely effect of rotation on the rise times of thin flux
tubes in low-mass stars, using numerical simulations within the thin flux
tube approximation.

\section{ASSESSING FIELD EVOLUTION USING SIMULATIONS}

Our analytical estimates of the buoyant rise of thin flux tubes in \S 2 are fairly
simplistic, as they assume an unchanging tube rising passively under the
influence of buoyancy.  In reality, even if a thin flux tube is produced in
the star, it will evolve as it rises under the action of magnetic buoyancy,
and in general we expect this to alter its rise time somewhat.  We aim in
this section to assess whether the broad conclusions from our earlier
analytical estimates are in keeping with somewhat more sophisticated models: namely,
whether thin flux tubes always rise more slowly than large ones, and
whether the rise time is of the same order as our earlier estimates.  

\subsection{Computational approach}

To numerically model the rise of isolated, fibril magnetic fields through a
0.3 solar mass star, we adopt the thin flux tube approximation
\citep[e.g.][]{defouw_1976,roberts_1978,spruit_1981a,choud_1987}.  The thin
flux tube approximation describes the 1D reduced MHD equations along the
flux tube axis subject to magnetic tension, buoyancy, Coriolis force, and
the drag force, with the equations we solve discussed in detail in
\citet{fan_1993}.  The flux tubes we study here rise through a quiescent
convective envelope with thermodynamic and stratification quantities taken
from a 1D model of a 0.3 solar-mass M-dwarf as described above.  

Following \citet{weber_2011}, our simulations start with toroidal magnetic
flux rings in mechanical force equilibrium and neutral buoyancy.  As in other thin flux tube models, each simulation explicitly considers only a single isolated tube. To ensure
initial neutral buoyancy, the internal temperature of the flux tube is
reduced compared to the external temperature.  There has been some debate
regarding the proper initial conditions for thin flux tube simulations,
with mechanical equilibrium generally favored over temperature balance in the solar context 
\citep[see e.g.][]{caligari_1998}.  In order to facilitate a magnetic
buoyancy instability, the flux tube is perturbed with small undular motions
consisting of a superposition of Fourier modes with azimuthal order from
$m=0$ to $m=8$ with random phase relations.  These perturbations mimic
those that would be provided by convective flows, however they are much
smaller in amplitude, on the order of a few cm s$^{-1}$ in the radial
direction.  Note that in general, instability need not occur first for m=0 (axisymmetric) modes, so portions of the flux tube may begin to rise before others. Each flux tube evolves adiabatically and is initiated with a
latitude of 10$^{\circ}$ above the equator at a radial distance of
$r_{0}=0.5R$ (see Tube A in Fig. \ref{fig:riseimg}).

A constraint of the thin flux tube approximation requires that $a_{0}/H_{p}
\le 0.1$ in the region where the flux tube is initiated.  At $0.5R$,
$H_{p}=$ 3.5\e{9} cm in our stellar structure model.  Therefore, we cannot
perform simulations where $a_{0}$ is in excess of $10^{8}$ cm.
Furthermore, the ratio of $a_{0}/H_{p}$ ought to remain small throughout
the entire computational domain, ideally less than 1-2.  As a result, we
stop our simulations once the flux tube apex has reached $0.95R$ (see Tube C in
Fig. \ref{fig:riseimg}), operating under the assumption that the rise time
through the remaining $0.05R$ is negligible compared to the total rise.  In
our simulations, flux tubes of small cross-sectional radius $a_{0} \leq
10^4$ cm as well as most flux tubes with $B_{0}=10^{4}$ G suffer from a
scenario where the magnetic field strength at the flux tube apex either
drops to zero, or weakens drastically in the upper convection zone to a few
hundred G, most likely unable to survive the remaining $0.05R$.  Such a
scenario is discussed in \citet{moreno_1995} in the solar context, where it
shown that flux tubes of near equipartition field strengths with small
cross-sectional radii exhibit a sudden catastrophic expansion and weakening
of the magnetic field at the flux tube apex.  For these reasons, we limit
our study in this Section to flux tubes of $B_{0}=10^{5} - 10^{7}$ G with
$a_{0}=10^{5} - 10^{8}$ cm.  The magnetic flux of each flux tube remains
constant such that $\Phi=B(\pi a^{2})$, resulting in flux values ranging
from 3.14\e{15} Mx - 3.14\e{23} Mx for our choice of $B_{0}$ and $a_{0}$.
In addition, the simulations are also carried out for stellar rotation
rates of 1, 3, 10, and 40 times that of the Sun, with $\Omega_{\odot}=$
2.7\e{-6} rad s$^{-1}$.

\subsection{Rise times}

\begin{figure}
\plotone{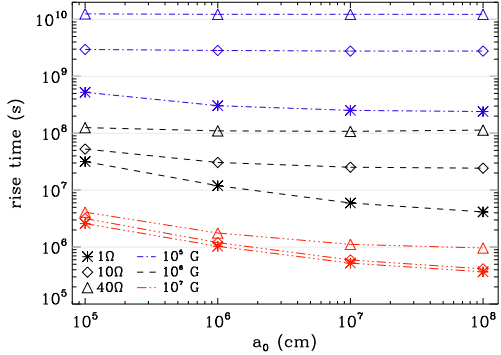}
\caption{Rise times in thin flux tube simulations as a function of initial cross-sectional radius $a_{0}$ for various $B_{0}$ and $\Omega_{0}$.  For each particular $B_{0}$, the rise time tends to increase as $a_{0}$ decreases.  However, rapid rotation normalizes this variation especially for smaller $B_{0}$.}
\label{fig:risetimes_tft}
\end{figure}

In Figure \ref{fig:risetimes_tft}, we plot (for models with
$B_{0}=10^{6} - 10^{7}$ G) the total time elapsed from the
beginning of the simulation to a time when some portion of the flux tube
has reached the simulation upper boundary.  For each particular $B_{0}$, in
accord with the analytical estimates of Section 2, the rise time tends to
increase as the cross-sectional radius $a_{0}$ decreases.  This is shown
most prominently in Figure \ref{fig:risetimes_tft} for a rotation rate of
1$\Omega_{\odot}$, and for $10^{7}$ G flux tubes at all rotation rates
considered.  Increasing the rotation rate also increases the rise time by
suppressing the growth rate of the magnetic buoyancy instability \citep[see
  e.g.][]{gilman_1970,schuessler_1996}.

\begin{figure}
\epsscale{0.8}
\plotone{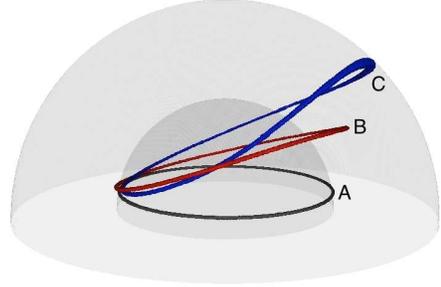}
\caption{Depiction of a $B_{0}=10^{6}$ G, $a_{0}=10^{8}$ cm flux tube rotating at $10\Omega_{0}$ for snapshots at three different times in its evolution.  Tube A shows the initial position of the flux tube at $r_{0}=0.5R$, with a darker inner hemisphere mapping out a surface of radius $r_{0}$.  Tube B and Tube C have apices that reach 0.75$R$ and 0.95$R$, respectively.  The simulation upper boundary at 0.95$R$ is shown by the outer gray hemisphere.  Rapid rotation forces the deflection of the tube apex to 47$^{\circ}$.  Each tube segment has been given a 3D extent according to the local cross-sectional radius, and the tube evolution is shown in a reference frame co-rotating with the star.}
\label{fig:riseimg}
\end{figure}

Rapid rotation also reduces the variation in rise times, effectively normalizing
it for smaller $B_{0}$.  The Coriolis force increases in magnitude relative
to the buoyancy and drag forces as the rotation rate is increased,
overwhelming the contribution to the rise time from the size of the tube
cross-sectional radius.  Furthermore, the increasing magnitude of the
Coriolis force relative to the buoyancy force results in a poleward
deflection of the rising loop \citep[in the rapidly rotating context, see
  e.g.][]{schuessler_1992,deluca_1997}.  An example of this scenario is
depicted in Figure \ref{fig:riseimg} for a $B_{0}=10^{6}$ G, $a_{0}=10^8$
cm flux tube rotating at $10\Omega_{\odot}$.  A non-axisymmetric m=1 unstable mode develops due to the small
amplitude perturbations applied to the initially stable tube. As the flux tube apex rises,
conservation of angular momentum drives a counter-rotating flow of plasma
elements along the tube which in turn induces an inward directed (toward
the rotation axis) Coriolis force.  This opposes the outward directed (away
from rotation axis) component of the buoyancy force. However, the Coriolis
force cannot balance the poleward component of the buoyancy force, and the
apex subsequently rises parallel to the rotation axis.  The degree to which
the tube will be deflected from radial motion depends on the relative
magnitude of the Coriolis force to the buoyancy force.  In Figure
\ref{fig:riseimg}, the apex reaches 47$^{\circ}$ at the simulation upper
boundary.  For comparison, given the initial position of our simulated flux
tubes, truly parallel motion to the rotation axis results in an emergence
latitude at $0.95R$ of $\sim$$58^{\circ}$.  Poleward deflection of the
rising tube also increases the distance the apex must traverse to reach the
surface, thereby increasing the rise time as well.  Flux tubes of $10^{7}$
G only begin to show moderate deflection of $\sim$$10^{\circ}$ poleward at
$40\Omega_{\odot}$ for the thickest tube ($a_{0}=10^{8}$ cm).  Due to their
large magnetic field strength, the magnitude of the Coriolis force is
always much less than the buoyancy force, even at high rotation rates.

The simulated rise times in Figure \ref{fig:risetimes_tft} are fairly close to
the analytical estimates in Section 2.  For example, for extreme fields 
of $10^{7}$ G, Equation (3) gives a rise time of about 8\e{6} s for a flux
tube of $a_{0}=10^{5}$ cm.  In our simulations, a flux tube with those
parameters rises about a factor of two faster than this if $\Omega$ is 40 times the
solar rate, and a factor of 3-4 faster if $\Omega/\Omega_{\odot}$ is
between 1 and 10.  Placing $r_{0}$ at $0.25R$ or the origin  would
increase the rise times given in Figure \ref{fig:risetimes_tft} somewhat.  A
smaller $r_{0}$ also increases the radius of curvature of the flux tube,
thereby increasing the magnetic tension and decelerating the buoyant rise
of the flux tube apex.  Including the effects of convective flows and
radiative diffusion on the motion of thin flux tubes in our simulations
will also have an impact on rise times; we expect this would be minimal for $10^{7}$ G flux
tubes owing to their extreme buoyancy, but could be more
pronounced at lower field strengths.  Though clearly
these differences would affect our maximum field estimates in Section 2 and
Section 3, the difference appears to us to be small in comparison to the
rather large theoretical uncertainties that underlie this whole subject.
We defer further discussion of the dynamics of rising flux tubes in fully convective stars, and their emergent
properties at the surface, to later work.

\section{OHMIC DISSIPATION AND HEATING}

Though the interiors of low-mass stars are highly ionized, they are not
perfect conductors. The currents sustained in the plasma must therefore
undergo Ohmic dissipation, with an associated heating (energy per unit time
per unit volume) $=j^2 / \sigma = 4 \pi \eta j^2/c^2$. Here $\eta$ is the
magnetic diffusivity (with units of cm$^2$ s$^{-1}$), related to the
conductivity of the medium by $\eta = c^2/(4 \pi \sigma)$.  Recall that the
current density ${\bf j}=(c/4\pi) (\nab\times{\bf B})$.

In other contexts, several authors have noted that for some magnetic field
strengths and morphologies, this Ohmic dissipation may represent a
significant heating source. For example,
\citet{liu_goldreich_stevenson2008} 
argued that this provided a powerful constraint on the depth of zonal winds
in Jupiter and Saturn: If the winds extend into the interior, they would
tend to stretch the observed surface poloidal fields into interior
azimuthal fields; if the winds extend too deeply, the heating associated
with the dissipation of these fields could in some cases exceed the
planet's luminosity.  Later, \citet{batygin_stevenson2010} invoked Ohmic
dissipation as a mechanism for inflating close-in extrasolar planets:
assuming reasonable profiles for the zonal winds and surface magnetism of
these objects, they argued that the associated Ohmic heating could be large
enough (and deposited deep enough) to lead to anomalously large radii.
Several other papers have investigated variants of this scenario in some detail,
assuming different models for the winds, conductivities, and magnetic
fields of these objects \citep[e.g.,][]{huang_cumming2012, wu_lithwick2013} and more recently simulating the atmospheric winds and magnetic
fields with varying levels of sophistication \citep{rauscher_menou2013,
  rogers_showman2014, rogers_komacek2014}.

In this section, we investigate the Ohmic dissipation associated with the
field strengths and morphologies examined in \S 2.  We give general
estimates of the current densities associated with these fields, and
briefly note a few firm limits on the currents associated with any field
distribution that is actively maintained against decay.  Employing
plausible conductivity profiles for the interiors of these objects, we
then calculate the power from Ohmic dissipation for fields of varying
strength and geometry.  Some combinations of field strength and topology
lead to very large rates of Ohmic heating, and we argue that this sets a
complementary limit on the strength of fields in the interiors of low-mass
stars.  

\subsection{Estimates and limits of current density}

The current density $j$ depends on both the strength and spatial scale of
the magnetism.  To order of magnitude, for a field varying on spatial scale
$a$, $j \sim cB/a$, so for fixed total field strength, smaller-scale fields
are associated with stronger currents.  We will generally employ this
simple estimate for our calculations below.

Before applying this estimate, we note that firm lower bounds can be
derived for the current density, but these will tend to underestimate the
magnitude of $j$ for small-scale fields.  Intuitively, it seems clear that
the current cannot be less than $j \sim cB/R$ with $R$ the radius of the
star, and it is straightforward to show that this is the case.  More
specifically, for divergence-free fields confined in a sphere of radius $R$
and matching to a decaying potential outside that sphere, one can prove
that
\begin{equation}
  \int{|\nab\times{\bf B}|^{2} dV} \ge \frac{\pi^2}{R^2} \int{|B|^2 dV}
\end{equation}
\citep[e.g.,][]{jones2008}.  Another possible constraint comes from consideration
of the induction and momentum equation: the rate of change of magnetic
energy is related to the Poynting flux out of the volume and to the work done
by the fluid ($\propto {\bf j} \cdot {\bf v} \times {\bf B} = - {\bf v}
\cdot {\bf j} \times {\bf B}$), in principle providing a constraint on the
minimum $j$ for a given rate of induction.  (To see this, note that ${\bf j}
\cdot {\bf v} \times {\bf B}$ is bounded by the product of $u_{\rm max}$ and
the square root of volume integrals over $j^2$ and $B^2$; see, e.g.,
Childress 1969, where a similar line of reasoning is used to provide a bound on
the minimum magnetic Reynolds number needed for dynamo growth.)  Related
arguments have also been used in an effort to infer, for example, the Ohmic
dissipation in Earth's core \citep[e.g.,][]{roberts_jones_calderwood2003}. In
practice these estimates typically provide only lower bounds, and we have
not found them to be particularly useful for constraining the current
density except in special circumstances.  Ultimately, any realistic field
distribution will likely possess fields over a range of spatial scales, but
it is always possible to find a characteristic $a$ such that the current
density, averaged over some volume, is $j \sim cB/(4 \pi a)$.  This need
not, and in general will not, correspond to the smallest or largest scales
physically present in the system.

\subsection{Conductivity profiles}

\begin{figure}
\center
\plotone{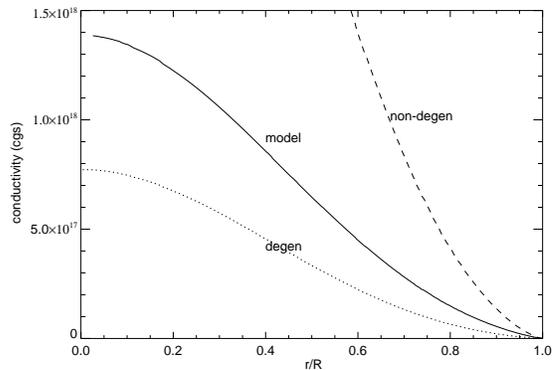}
\caption{Estimates of the radial variation of conductivity in the interior of a low-mass star (in
  cgs units).  Shown are the values adopted in this work (solid line), together with estimates for purely degenerate
  and non-degenerate matter (see text). \label{fig:conductivity}}
\end{figure}

For specificity, we primarily consider the interior of a 0.3 solar mass
M-dwarf, whose temperature, density, and pressure are as in previous
sections given by a 1-D
stellar model \citep{cb1997}.  These models calculate the
conductivity using the methods of \citet{potekhin1999, potekhin_etal1999}. Although 
 initially developed for neutron star envelopes and white dwarf
cores, these recover previous calculations
\citep{hubbard_lampe1969,itoh_etal1983, mitake_ichimaru_itoh1984,
  brassard_fontaine1994} at lower
densities and cover the range of temperatures and densities characteristic
of low-mass stars and brown dwarf interiors. They have been used for
instance in \citet{chabrier_etal2000} to calculate the conductive opacities in
old and massive brown dwarfs. They are presently the most detailed and
accurate conductivity calculations for conditions ranging from neutron
stars to jovian planets. The conductivity
profile is shown as the solid line in Figure \ref{fig:conductivity}.  In the regime considered here, the conductivity can be taken to be independent of the magnetic field strength and morphology.

For ease of comparison with other work, we note that the conductivity
profile employed here is not too different from that derived by other
methods.  A basic rule of thumb for electron conduction is that the
magnetic diffusivity (in cm$^{2}$ s$^{-1}$) is $10^4 T^{-1.5}$, with $T$ in
millions of K.  \citet{huang_cumming2012} also give formulae for conductivity
in partially degenerate matter, and we quote these here for convenience.
In general we have $\sigma = n_e e^2 / m_e \nu$, with $\nu$ the collision
frequency and $n_e$ the electron number density.  If the plasma is ionized
enough for electron-proton collisions to dominate, $\nu \approx 4 e^4 m_e
\Lambda / (3 \pi \hbar^2) \approx 1.8 \times 10^{16}$ s$^{-1}$ in the fully
degenerate limit, or $\nu \approx 6.4 \times 10^{23}$ s$^{-1}$ $\rho Y_e
T^{-3/2}$ if not fully degenerate.  Purely for comparison purposes, we show both the fully degenerate and
non-degenerate conductivity curves in Figure \ref{fig:conductivity} as well
(dashed and dotted lines).  The conductivity profile adopted here is
intermediate between these two extremes.

A typical value of $\sigma$ in the deep interior (namely, the mean value
over the inner $1.5 \times 10^9$ cm) is about $8 \times 10^{17}$ (cgs),
implying a magnetic diffusivity of about 80 cm$^2$ s$^{-1}$.  The ``rule of
thumb'' calculation above gives similar values.  Assuming the plasma is
fully degenerate would result in a typical conductivity about 0.16 times
the non-degenerate value, or equivalently diffusivities that are about a
factor of 6 larger.

\subsection{Ohmic dissipation for proposed field distributions}

The energy per unit volume per unit time from Ohmic dissipation is given by
the square of the current density divided by the conductivity.  The total
power associated with Ohmic dissipation in a volume extending from the
origin to radius $R$ is
\begin{equation}
  L_{\rm dissip} = \int_{0}^{R} 4 \pi r^2 \frac{j(r)^2}{\sigma(r)} \,dr
\end{equation}
where all symbols take their usual meanings.  For any given field distribution and characteristic scale $a$, it is
then straightforward to calculate the total Ohmic dissipation.

In this section we briefly investigate the dissipation associated with some
of the strong field distributions examined in previous work (FC2014, MM).
These published models are characterized only by a characteristic field
strength at every depth; the morphology of the field is left unspecified.
Clearly the characteristic spatial size of the field $a$ has a significant
impact on the dissipation: the total power from dissipation scales like
$(B/a)^2$, so large-scale fields lose less energy (per unit time) to
dissipation than small-scale ones.

In one of the models considered extensively in FC2014 (their ``Gaussian''
model), the magnetic field strength is assumed to vary with radius as
\begin{equation}
  B_{\rm FC14} = B_{\rm max} e^{-0.5((r_t - r)/\sigma)^2}
\end{equation}
where $r_t$ and $\sigma$ are 0.15 and 0.10 times the maximum radius
respectively, and $B_{\rm max} = 4.0 \times 10^7$ G.  We
assume the simple model for current density above ($j \sim cB/4 \pi a$),
and consider three representative values of $a$, namely $10^6$, $10^7$, and $10^8$ cm.  It is worth noting that
at this extreme field strength, even these (comparatively small scale) fields would not actually be stable
against magnetic buoyancy according to the most stringent of our limits
above; if the field were composed entirely of thin flux tubes, they would
need to have a typical radius $a$ below $10^5$ cm  (which would make $j^2$
about 100 times larger than our largest estimates here) to be stable.

\begin{figure}
  \epsscale{1.2}
\center
\plotone{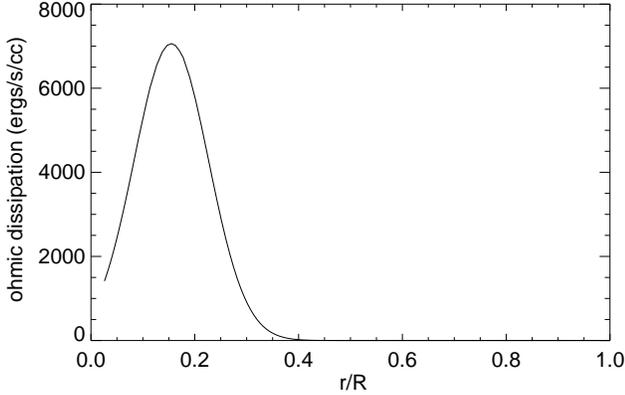}
\caption{Heating from Ohmic dissipation (energy per unit volume per unit
  time) dissipated for the FC2014 field if $a=10^6$ cm and the conductivity
profile is as assumed above.\label{fig:ohmicfc2014}}
\end{figure}

\begin{figure}
  \epsscale{1.2}
\center
\plotone{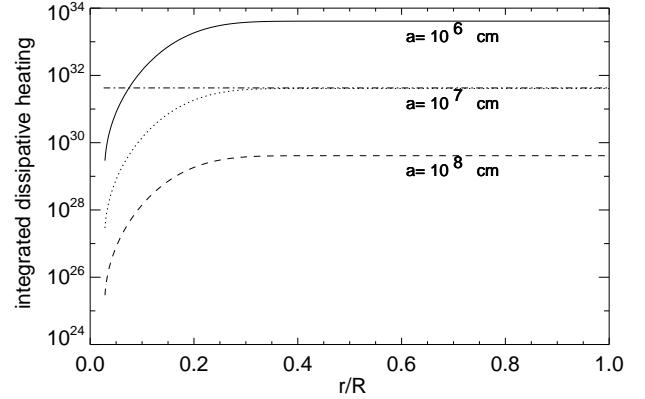}
\caption{Total integrated luminosity from Ohmic dissipation in the same
  model as a function of radius, for three different values of
  characteristic dissipative scale $a$, and compared to the total stellar
  luminosity (shown as dashed line). \label{fig:ohmiclumfc2014}}
\end{figure}

Figure \ref{fig:ohmicfc2014} considers the Ohmic dissipation at each radius for this model,
assuming the conductivity is given by the model above, and
Figure \ref{fig:ohmiclumfc2014} calculates the integrated power from Ohmic dissipation ($L_{\rm
  dissip}$ above), with the total luminosity of the star overplotted as a
horizontal dashed line.  It is clear that the combination of very strong fields (as
required in the FC2014 model to explain inflated radii) and fairly small
characteristic lengthscales would result in Ohmic dissipation exceeding the
luminosity of the entire star.  Larger values of $a$ lead to less
dissipation, with the heating scaling as $a^{-2}$.  Thus, while adopting the
FC2014 peak field strength and values of $a=10^6$ cm leads to dissipative
luminosities that vastly exceed the stellar luminosity, and taking $a=10^7$ cm
cm yields dissipation of the same order as $L_{*}$, values of $a \ge 10^8$
cm at the same field strength would lead to fairly negligible heating.

\subsection{Minimum spatial scales consistent with Ohmic constraints}

The 1-D stellar models in general use today include no explicit allowance
for Ohmic heating in the interior, so situations in which the integrated
Ohmic dissipation approaches the stellar luminosity represent a
possible contradiction.  We can therefore use the constraint that the total
dissipation not exceed $f L$, where $f$ is a factor less than 1, to
constrain the current density within the star.  Furthermore, although the
total dissipation in some convective systems may exceed the total
luminosity without violating the laws of thermodynamics (see
\citealt{hewitt_mckenzie_weiss1975}, and explicit numerical examples in
\citealt{jones_kuzanyan2009, viallet_etal2013}), this is only the case when the (thermal)
stratification is strong.  For the gradually-declining temperatures in the
interior of a low-mass star, $f$ must be below unity.  (Formally, in many
cases it is bounded by a value of order $r/H_t$, with $H_t$ the thermal
scale height, which in turn implies $f$ less than unity for the deepest portion of the 
interior.) For a given profile of $B$ (as in the FC2014 profiles considered
above), and assuming the current density is related by the simple scaling
relations above to the characteristic spatial scale of the field, we can
then derive a {\sl minimum} characteristic scale for the field.
Smaller-scale fields (at fixed overall field strength) would result in too
much dissipation.  The resulting estimate is
\begin{equation} \label{eqn:amin}
  a_{\rm min} = \left[ \frac{ \int_0^{R} 4 \pi r^2 \left(\frac{c B(r)}{4
        \pi} \right)^2 \frac{1}{\sigma (r)} \,dr}{f L} \right]^{1/2}
\end{equation}

We calculate $a_{\rm min}$ for specific field distributions from FC2014
below.  First, though, we note a few cases in which this equation has a
particularly simple interpretation.  If the dissipation $j^2 /\sigma$ does
not depend on position within the star, then it is straightforward to show
that our expression for $a_{\rm min}$ is equivalent to
\begin{equation}
  \label{eqn:aminequipart}
  a_{\rm min}^2 \sim \bar{\eta} R^3 \frac{1}{f L} \bar{B^2},
\end{equation}
where $\bar{\eta}$ is the magnetic diffusivity averaged over the volume
and $\bar{B}$ is the average field strength.  This criterion is more
transparently written as
\begin{equation}
  \frac{E_{\rm mag}}{\tau_{\eta}(a_{\rm min})} \lesssim fL
\end{equation}
where $E_{\rm mag} = (4 \pi R^3/3)(B^2/8\pi)$ is the total magnetic energy
in the volume and $\tau_{\eta}(a_{\rm min})$ is the magnetic diffusion time
for fields on scale $a_{\rm min}$.  If $a$ is reduced below $a_{\rm min}$,
the characteristic diffusion time becomes shorter, so the power dissipated
per unit time is larger; we require that it be smaller than the luminosity
of the star (or if $f \neq 1$, some fraction thereof).

For the specific 1-D stellar model considered here, with average Ohmic
diffusivities as quoted above, this requirement translates to a linear
relationship between $a_{\rm min}$ and the average field $B$, namely
$a_{\rm min} \approx \sqrt{6/f} \bar{B}$ cm G$^{-1}$, where the total power from Ohmic
dissipation is not to exceed $fL$.  Although values of $f$ approaching or exceeding 
unity are possible in some cases, we will take $f=0.4$ as a reasonable limit for the deep interior: above
this, we expect the structure of the star would be significantly affected
by the Ohmic dissipation, leading to changes that are at least as great as
those purportedly arising from modifications to the convective heat
transport.  For this value of $f$, $a_{\rm min} \approx 3.9 B$ cm G$^{-1}$,
so values of $B \approx 10^6$ G require that the field be structured
predominantly on scales larger than about $4 \times 10^6$ cm.

The crude assumption that $j^2/\sigma$ is independent of depth is not as
unrealistic as it might at first appear.  Suppose the field strength at
every depth is a fixed multiple of the equipartition field strength, $B
\propto \sqrt(\rho) v_c$, and that the convective velocity does not vary
too much with depth, so $B^2 \propto \rho$.  The density and temperature structure of a fully
convective star is reasonably well approximated by a polytrope with index
$n=1.5$, in which case the density and temperature follow relations of the
form $\rho = \rho_c (\theta(\xi))^{1.5}$, $T=T_c \theta(\xi)$, with
$\xi$ a dimensionless radius $r = r_n \xi$ (in turn involving a scale
length $r_n^2 = (n+1)P_c/(4 \pi G \rho_c^2)$).  Recall that the
conductivity in the non-degenerate case is $\sigma \propto T^{3/2}$, which
in the n=1.5 polytropic case implies $\sigma(r) \propto   \theta^{3/2} \propto
\rho(r)$.  Hence for the scaled-equipartition field, the ratio $B^2/\sigma
\approx \rho(r)/\rho(r)$ is approximately constant.

It is worth reiterating that for more realistic field distributions containing a
range of spatial scales, with $B$ in general a function of spatial scale,
one could still define an $a_{\rm min}$ consistent with the requirement
that Ohmic dissipation not exceed a given luminosity.  In general this
estimate would correspond neither to the largest scales present in the
system nor the smallest, but to a characteristic scale intermediate between
the two.  For the conductivity profiles and field strengths considered here, the microphysical
dissipation scale, taken as the scale $l$ at which the magnetic Reynolds
number $Rm = u(l) l/\eta \approx
1$, is much smaller than $a_{\rm min}$ for any plausible variation of $u$
with length $l$: e.g., for $u(l) = u(R) (l/R)^{\alpha}$ as before, the
dissipative scale is given by
\begin{equation}
 l_{\eta} \approx \left(\frac{\eta R^{\alpha}}{u(R)} \right)^{\frac{1}{1 +
     \alpha}}.
\end{equation}
If $\alpha = 1/3$, this reduces to the familiar rule that $l_{\eta} \propto
Rm^{-3/4}$, whereas if $\alpha=1$ it is equivalent to the statement that
the dissipation time at $l_{\eta}$ is equal to the large-scale dynamical
time $R/u(R)$. In both cases $l_{\eta} \ll a_{\rm min}$.  

In what follows, we will generally employ both the simplest estimate of
$a_{\rm min}$ above (i.e., assuming $j^2/\sigma$ is approximately constant) and a
slightly more complex calculation in which we have numerically integrated
equation~\ref{eqn:amin} for specific field profiles from FC2014 (assuming
the conductivity is given by the 1-D model as above).  Both estimates of
$a_{\rm min}$ scale in the same way with $B$, but they differ by constant
factors of order unity.  In both cases, because $a_{\rm min}$ increases
with $B$, conflict with the requirements of \S 2 that the field be {\sl
  larger} than some given scale to avoid rapid losses due to magnetic
buoyancy instability, are in principle possible.  We explore this conflict
in the next section.

\section{COMBINED LIMITS ON FIELD STRENGTH}

We have so far highlighted a few of the difficulties involved in finding extremely
strong magnetic field configurations that could persist for indefinite
intervals.   In \S 2 we reviewed a variety of work on magnetic buoyancy
instabilities in stellar interiors.  We have not analyzed arbitrarily
complex field distributions, but have considered two extreme cases:
isolated flux tubes initially in equilibrium with their field-free
surroundings, and smooth spatially-varying distributions of magnetism.
Both the strength of the field and its spatial structure modify the growth
of these instabilities.  Fields as strong as those considered in some 
models are likely only stable against buoyancy (or have
buoyant rise times shorter than plausible regeneration timescales for the
magnetism) if the field is predominantly structured on very small scales.
An equivalent statement is that at any fixed field strength, fields larger
than some maximum spatial scale $a_{\rm max}$ are unstable; smaller-scale
fields might conceivably persist.  At fields of order the equipartition strength, this
maximum allowable spatial scale extends to the largest possible scale in
the system (the stellar radius), but at very strong field strengths only
small-scale fields are consistent with the constraint of buoyancy.

But associated with the small-scale fields necessitated by buoyancy are
intense currents.  In \S 4 we argued that the Ohmic heating associated with
strong, small-scale fields would greatly exceed the stellar luminosity in some
cases.  Equivalently, because the Ohmic heating associated with a field of
magnitude $B$ and average spatial scale $a$ scales with $B/a$, at fixed
field strength there is a minimum spatial scale $a_{\rm min}$ consistent
with the assumed stellar luminosity: fields structured predominantly on
smaller scales dissipate too much energy.

\begin{figure}
  \epsscale{1.2}
  \center
  \plotone{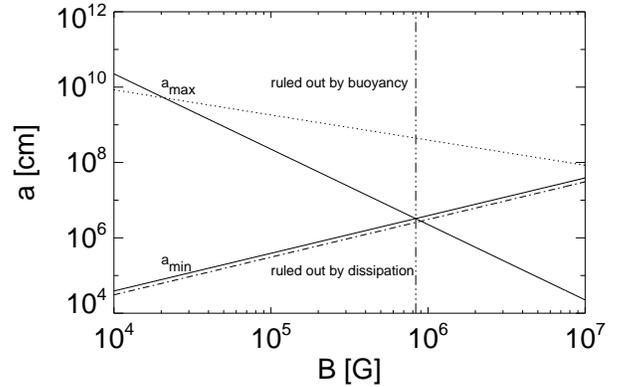}
  \caption{Characteristic spatial scales of magnetism that are ruled out by Ohmic
    dissipation and magnetic buoyancy.  At given field strength,
    configurations consisting primarily of very large scale fields are
    ruled out by magnetic buoyancy, while very small-scale fields are
    inconsistent with the Ohmic dissipation constraint.  Beyond a maximum
    field strength $B_{\rm crit}$, denoted by a dashed vertical line for a
    particular choice of models, no spatial scales of the field are consistent with both
    constraints. \label{fig:combinedconstraints}}
\end{figure}

These constraints can be combined to give an approximate limit on the
maximum possible field strength in the stellar interior.  In Figure
\ref{fig:combinedconstraints} we show the estimates of \S2, which give
$a_{\rm max}$ for a collection of flux tubes at any given field strength
$B$, together with the calculation from \S4 of the {\sl minimum}
characteristic scale $a_{\rm min}$ consistent with the constraints of Ohmic
dissipation.  The solid lines show the analytical estimates of equations
\ref{eqn:aminequipart} and \ref{eqn:amaxslow} (employing a typical depth
$r=0.25R$ for the estimate of the rise time), which we take as our most representative calculations.  At low
field strengths (of order the equipartition value), a wide range of
characteristic scales are compatible with both constraints.  At higher
field strengths the window of allowable field strengths narrows, and
eventually the two lines $a_{\rm min}$ and $a_{\rm max}$ cross, indicating
that no characteristic field scale can simultaneously meet both
constraints.  We regard this intersection as an upper limit on the
achievable field strength: above it, any field that is stable to magnetic
buoyancy would result in too much Ohmic heating.

The actual value of $B_{\rm max}$ naturally depends on what estimates are
adopted for $a_{\rm min}$ and $a_{\rm max}$, but an upper limit (i.e., an
intersection of $a_{\rm min}$ and $a_{\rm max}$) exists for all the models
we have considered here.  For the specific model of $a_{\rm max}$
considered in equation \ref{eqn:amaxslow}, which estimates the maximum
field that could be pumped downward against magnetic buoyancy (or,
equivalently, the largest field that could be regenerated by slow,
large-scale eddies), and for the estimate of $a_{\rm min}$ from equation
\ref{eqn:aminequipart} (which assumes the dissipation $j^2/\sigma$ is
independent of depth, and allows Ohmic dissipation up to 40 percent of the
stellar luminosity), the resulting maximum field strength is about 800 kG.
More generous assumptions about the effectiveness with which flows could
rebuild magnetism as it is emptied out by magnetic buoyancy, as represented
for instance by the ``fast eddy/shear'' model of field generation and the
upper (dotted) $a_{\rm max}$ line in Figure \ref{fig:combinedconstraints},
could lead to larger $B_{\rm max}$.  Adopting the intermediate ``cascade'' model for field
regeneration (with a velocity dependent on scale $l$ as $v \propto
l^{\alpha}$, $0 < \alpha < 1$) leads to a maximum field estimate intermediate between these
two. We suspect that the strictest of these limits is likely to be the most
robust, in part because it also reflects a limit on the
maximum field that could persist for some time against magnetic buoyancy
{\sl without} ongoing rapid regeneration by dynamo action.  (Of course such a
field would inexorably decay from Ohmic dissipation, but this process is
slow when compared to the timescales associated with magnetic buoyancy or
convection.)

For comparison, we have also plotted (as the dashed-dot line) $a_{\rm min}$ calculated for a field
whose variation with radius follows the ``Gaussian'' profile from FC2014 as
described above, and which attains a {\sl maximum} value given by the
x-axis.  At each point we have numerically calculated the integral over
$j(r)^2/\sigma(r)$, still assuming that $j(r) = cB(r)/(4 \pi a)$, and
allowing $f=1$.  The variation with $B$ is the same, but the curve is
offset slightly with respect to our simple model (and slightly different
value of $f$), so the maximum field strength consistent with both buoyancy
and dissipation is slightly higher (about 1 MG). The conclusions remain otherwise the same as in the simpler model.

An explicit expression for $B_{\rm max}$ follows from setting $a_{\rm min}
\approx a_{\rm max}$.  In the specific case of ``slow'' eddies represented
by the solid line in Figure (\ref{fig:combinedconstraints}), and taking $H_p \approx R$ in the interior
as in equation \ref{eqn:amaxequipart}, we have that
\begin{equation}
  B_{\rm max}^3 \approx B_{\rm eq}^2 \sqrt{\frac{fL}{\eta R}}
\end{equation}
where as before $B_{\rm eq}$ is an estimate of the field in equipartition
with the convective kinetic energy.  The term under the square root can be
understood intuitively as the extraordinarily strong field that would
satisfy $a_{\rm min} = R$: i.e., the field for which the Ohmic dissipation
associated with dissipation on the largest possible scale $R$ would still exceed the luminosity of the
system,
\begin{equation}
B_{\eta} \approx \sqrt{\frac{fL}{R \eta}}.
\end{equation}
For the model considered here, $B_{\eta}$ is in excess of $10^9$ G, and
$B_{\rm eq} \sim 10^4$ G, yielding the quoted $B_{\rm max}$ of order $8
\times 10^5$ G.

We emphasize that our estimates here refer to the maximum mean field
that could persist, in the same way that our estimates of scale 
correspond to a single spatial scale that characterizes the overall field
distribution.  This is convenient for comparison with prior models of the
interior magnetism in low-mass stars (e.g., the models of FC2014 and MM),
which are likewise characterized by a single field strength at each depth.
More realistic field configurations will of course contain a range of field
strengths and morphologies, which may exceed $B_{\rm max}$ in some
locations and contain structures smaller than $a_{\min}$ without violating
in a global sense any of the constraints explored in this paper.  But it is always
possible to define a mean field and a characteristic spatial scale, and
these must obey some variant of the constraints noted here.

\section{DISCUSSION: TOWARDS EXTENSIONS TO OTHER MASSES}

Our discussion thus far has concentrated on the constraints provided by
buoyancy and dissipation in the interior of a fully convective 0.3
solar-mass M-dwarf.  Here, we present a very simple estimate of how these
constraints might scale to other masses, while retaining the assumption
that the interior of the star is fully convective.  As noted above, the
extension to cases with a stably stratified core is somewhat more complex,
and considerably more uncertain, so we defer it to later work.

We concluded in \S5 that the maximum field strength consistent with the
constraints of both Ohmic dissipation and magnetic buoyancy is, to order of
magnitude, $B_{\rm max}^3 \approx B_{\rm eq}^2 B_{\eta}$, with
$B_{\eta}=\sqrt{fL/(R \eta)}$ and $B_{\rm eq} = \sqrt{4 \rho}v_c$.  We also
assume that the convective velocity follows the mixing-length scaling, $v_c
\propto (L R/M)^{1/3}$.  Let us further assume a power-law relationship
between stellar mass and luminosity, $L \propto M^{\beta}$, and take mass
roughly proportional to stellar radius, $R \propto M$.  (This latter
assumption will break down in degenerate objects, but is reasonable for
objects on the hydrogen-burning main sequence.)  We further note that for
an object composed of ideal gas in hydrostatic equilibrium, the central
temperature is roughly constant in this mass regime \citep[e.g.,][]{cb1997}, implying
$\eta$ is in turn approximately constant.  Then $B_{\eta} \propto
\sqrt{M^{\beta}/M} = M^{(\beta -1)/2}$.  The equipartition field strength
scales like $B_{\rm eq} \propto M^{(\beta -3)/3}$, so some manipulation
yields
\begin{equation}
  B_{\rm max} \propto M^{\frac{7 \beta - 15}{18}}.
\end{equation}
In 1-D models \citep{cb1997}, $\beta \approx 5$ for masses between 0.5 and 1 solar masses,
and $\beta \approx 2.4$ between 0.1 and 0.5 solar masses.  (The eventual flattening of the mass-luminosity
relationship in these models is due primarily to the formation of $H_2$ and the onset of
convection in the atmosphere, rather than to deep interior properties.)
Hence $B_{\rm max}$ would be expected to vary only weakly with mass in the
very low mass star/brown dwarf regime (for $\beta=2.4$, $B_{\rm max}
\propto M^{0.1}$).

\section{CONCLUSIONS AND PERSPECTIVES}

We have aimed to examine here whether the very strong interior magnetic
fields invoked in some models of low-mass stars, however initially
established, could be maintained indefinitely.  For the very simple field
configurations considered here, which consist primarily of collections of thin flux
tubes, we find that Ohmic dissipation and magnetic buoyancy combine to
yield interesting constraints on possible field strengths and
morphologies. We argued in \S2 that strong fields might avoid rapid rise from magnetic buoyancy
instability, or be regenerated faster than they rise, only if they are
structured on very small scales.  But the Ohmic dissipation associated with
such small-scale magnetism would in some cases exceed the luminosity of the
star (\S4). In \S5 we concluded that above a strength of order 1 MG, no
field can satisfy both constraints simultaneously, so we regard this as a
practical upper limit on the fields that could be maintained.

A principal limitation of our work is its reliance throughout on very
simple models of the magnetic field's spatial structure.  Our calculations
of the rise time of buoyant fields draw on the thin flux tube
approximation, which essentially asserts that the field is composed of
well-defined tubes characterized only by their strength and cross-sectional
area.  We argued that structures akin to these tubes might arise, for
example, from the breakup of smooth magnetic layers, which are also
provably subject to buoyancy instabilities. This approach is not
unreasonable, having been used to considerable effect in the solar
community for decades, and it has the great advantage that it allows us to
estimate quantities like the rise time analytically.  But it is still only
an approximation. In particular, our assumption that the rise of buoyant
flux is impeded essentially by drag, while also in line with many previous
authors, is likely a considerably simplified description of a more complex
process \citep[see, e.g.][]{hughes_falle1998}.  Assessing whether more complex
configurations of field are as susceptible to buoyant instability, and
likewise the rate at which such fields are really regenerated by dynamo
action, is probably not possible without recourse to more sophisticated
numerical simulations.

Likewise, our estimates of the Ohmic dissipation assume for simplicity that
the field can be characterized by only one spatial scale; in reality, any
dynamo-generated field will contain a range of spatial and temporal scales,
and the dissipative heating may involve more complex reconnection processes
than considered here \citep[e.g.,][]{lazarian_vishniac1999}.  We have also
ignored spatial anisotropies in the magnetism (or the flow), whether
induced by rotation or by strong fields themselves, even though these are
surely present \citep[see, e.g.,][]{oruba_dormy2014, davidson2013}.  But it
is always possible to define a characteristic scale for the field, and for
its dissipation, lying intermediate between the largest scales in the
system and the smallest.  Ultimately we think the dissipation, whatever its
distribution with scale, must still obey some form of the constraints
examined here.

In other ways, though, our estimates are very conservative.  In our
estimates of plausible regeneration times, for example, we have adopted an
essentially kinematic estimate (taking the growth time for the magnetism to
be of order the convective turnover time); in general, one would expect the
very strong fields examined here to exert considerable Lorentz feedbacks on
the flow, leading in turn to much slower regrowth than estimated here.
This would act to lower the maximum sustainable field strength.  

Perhaps the most robust conclusion from our work is that any model for the
interiors of low-mass stars that invokes very strong magnetic fields must
consider not just the strength of those fields but also their morphology.
Fields with different spatial distributions behave differently: if the
field is mostly on small scales, it might conceivably be regenerated faster
than buoyancy instabilities can act to remove it, whereas large scale
fields take longer to regenerate and are more susceptible to rapid buoyant
losses.  On the other hand, fields that vary on small scales are subject to
stronger Ohmic dissipation, and we have shown that this presents
significant constraints in the most extreme cases.

The full implications of this work for the structure of low-mass stars are
not yet clear.  Other authors (e.g., FC14) have indicated that
  magnetic fields of great strength would be required to appreciably
  inflate the radii of low-mass stars (within the context of a particular
  mixing-length model for the convective transport); FC14 in particular
  ultimately concluded that such fields were unlikely for a variety of
  reasons.  Our work partly echoes and complements theirs, by showing explicitly
  that strong fields on small spatial scales conflict with the combined
  constraints provided by buoyancy and Ohmic dissipation.  Despite the
many caveats noted above, we think that the basic constraint provided by
buoyancy and dissipation is likely to be robust, and hence that the very
strong fields examined in some previous models are not feasible.  Assuming
that the conflict between observed and predicted radii in these stars is
real, this suggests either that other phenomena are acting to ``inflate''
the stars, or that even weaker magnetism (coupled with rotation) can affect
the convective transport.  We intend to examine these possibilities in
future work.

\acknowledgements

We thank Isabelle Baraffe for providing structure models used in this
paper, and for many helpful discussions. This work was supported by the
European Research Council under ERC grant agreements no 337705 (CHASM) and
(FP7/2007-2013) no 247060, and by a Consolidated Grant from the UK STFC
(ST/J001627/1). We also thank Andrew West, who supported AM's work on
this project. Some of the calculations for this paper were performed on the
DiRAC Complexity machine, jointly funded by STFC and the Large Facilities
Capital Fund of BIS, and the University of Exeter supercomputer, a DiRAC
Facility jointly funded by STFC, the Large Facilities Capital Fund of BIS
and the University of Exeter.

\bibliographystyle{apj}
\bibliography{buoyancy_dissip_extracted.bib}

\end{document}